\documentclass[onecolumn, sort&compress, numbers]{els-mrw} 
\usepackage{amsmath,amssymb,amsfonts,amsthm,makeidx,graphicx}
\usepackage{xcolor}
\usepackage{txfonts}
\usepackage{helvet}
\usepackage{slashed}
\usepackage{bbm}
\usepackage{multirow}
\usepackage{ytableau,youngtab}

\newcommand{\Diff}[2]{{\rm d}^{#2} #1 \,} 

\newcommand{\e}{\mathrm{e}}

\newcommand{\<}{\langle}
\renewcommand{\>}{\rangle}
\newcommand{\vev}[1]{\langle #1 \rangle}
\newcommand{\ket}[1]{| #1 \rangle}

\usepackage{slashed} 
\newcommand{\mc}[1]{\mathcal{#1}}
\newcommand{\tr}{\mbox{tr}}
\usepackage{hyperref}

\usepackage{xcolor}

\begin{document}


\chapter{An introduction to gauge theories and group theory in particle physics}\label{chap1}

\author[1]{Hao-Lin Li}%
\author[2,3]{Hao Sun}%
\author[4]{Ming-Lei Xiao}%
\author[5,6,7]{Jiang-Hao Yu}%

%
\address[1]{School of Physics, Sun Yat-Sen University, Guangzhou 510275, P. R. China}
\address[2]{\orgname{Institute of High Energy Physics, Chinese Academy of Sciences}, \orgaddress{Beijing 100049, China}}
\address[3]{CAS Center for Excellence in Particle Physics, Beijing 100049, China}
\address[4]{School of Science, Sun Yat-Sen University, Shenzhen 518107, P. R. China}
\address[5]{Institute of Theoretical Physics, Chinese Academy of Sciences,    Beijing 100190, P. R. China}
\address[6]{School of Physical Sciences, University of Chinese Academy of Sciences,   Beijing 100049, P.R. China}
\address[7]{School of Fundamental Physics and Mathematical Sciences, Hangzhou Institute for Advanced Study, University of Chinese Academy of Sciences, Hangzhou 310024, China}

\articletag{}

\maketitle

\begin{abstract}[Abstract]

    In this review, the fundamental concepts of group theory and representation theory are introduced. Special emphasis is placed on the unitary irreducible representations of the $SU(N)$ Lie group, the Poincaré group, Little Group, discrete group, and their applications in particle physics. Based on the principle of local gauge symmetry, the construction of gauge-invariant Lagrangians and their quantization procedure are discussed. To address gauge redundancy, the modern on-shell amplitude approach is applied to gauge theories, demonstrating both conceptual and computational advantages. From the perspective of symmetry, the Standard Model is presented through the identification of its gauge symmetry, its anomaly-free matter content, and its global symmetries—including flavor symmetry, custodial symmetry, and baryon and lepton number conservation, etc.

\end{abstract}

\begin{keywords}
Group Theory \sep Gauge Theory \sep Poincaré group \sep Little Group \sep Symmetric Group \sep Discrete Group \sep Gauge Redundancy \sep BRST symmetry \sep Yang-Mills \sep On-shell Scattering Amplitude \sep Standard Model \sep Flavour Symmetry \sep Custodial Symmetry  \sep Gauge Anomaly \sep Grand Unification 
\end{keywords}


\begin{glossary}[Nomenclature]
	\begin{tabular}{@{}lp{34pc}@{}}
        QED &   Quantum Electrodynamics \\
        QCD &   Quantum Chromodynamics \\
        FP  &   Faddeev-Popov \\
        BRST &  Becchi-Rouet-Stora-Tyutin \\
        ST  &  Slavnov-Taylor \\
        GUT &  Grand Unified Theory \\
        rep-mat & Representation Matrix \\
        SM & Standard Model\\
        EWSSB & Electro-Weak Symmetry Spontaneous Breaking \\
        VEV & Vacuum Expectation Value \\
	\end{tabular}
\end{glossary}

\section*{Objectives}
In this chapter, the reader will:
\begin{itemize}
\item Understand the basic concepts in group theory and representation theory, and how they are applied in particle physics. In particular, the representation theory of the symmetric group, $SU(N)$ group, and Poincaré groups are detailed, including how to systematically construct their unitary irreducible representations, such as the Young diagram and Little group.
\item Grasp the principle of local gauge symmetry and its similarity to gravity, which necessitates gauge fields. Follow the quantization process, including the Faddeev-Popov procedure and the role of BRST symmetry in defining the physical Hilbert space. Clarify the vital distinction between gauge redundancy and physical symmetry and explain why gauge anomaly cancellation is mandatory. The modern on-shell amplitude approach is introduced, bypassing gauge redundancy by building the S-matrix directly from physical constraints.
\item Gain an understanding of the standard model from the viewpoint of group theory. The gauge group of the standard model is discussed with its anomaly-free matter content, the representations of the field, and the gauge interaction. The global symmetries, including the flavor symmetry, custodial symmetry, and baryon and lepton number conservation, etc, are highlighted.
\end{itemize}



\section{Introduction}\label{intro}


Symmetry is the foundational concept of modern physics, fundamental to our understanding of nature's laws. We observe that physical phenomena often remain unchanged under certain transformations—such as rotations in space, translations, or shifts in time. The powerful mathematical framework for classifying such transformations is group theory. For instance, the symmetries underpinning Einstein's special relativity—rotations and boosts in Minkowski spacetime—form the Poincaré group. The profound implication is that the very structure of spacetime itself dictates this symmetry. In quantum mechanics, this relationship deepens; a continuous symmetry transformation is represented by a linear, unitary operator, furnishing a representation of a Lie group. The Hermitian generators of these transformations correspond to physical observables, such as momentum and angular momentum, directly linking abstract symmetry to measurable quantities.

This conceptual journey, from symmetry as a descriptive property to a generative principle, forms the core narrative of twentieth-century particle physics. It is an intellectual odyssey that begins not in the subatomic realm, but in the fabric of spacetime. The theory of special relativity can itself be understood as the requirement that the laws of physics be invariant under the Poincaré group. This spacetime symmetry elegantly unified space and time, and its representations were later found to classify elementary particles by their masses and spins.

The indispensable link between symmetry and physical law was crystallized by Emmy Noether in 1918~\cite{Noether:1918zz}. Her eponymous theorem establishes an indelible connection between continuous global symmetries and conserved quantities. Translational symmetry in time yields energy conservation; in space, momentum conservation; and rotational symmetry, angular momentum conservation. Noether's theorem provided the foundational rationale for why symmetry was not a mere mathematical curiosity but a physical imperative—conservation laws are the observable fingerprints of hidden symmetries.

The pursuit of symmetry deepened with general relativity, where the principle of general covariance—invariance under all continuous coordinate transformations—became the cornerstone, describing gravity as the curvature of spacetime. It was in this fertile ground that the seed of the gauge principle was planted. In 1918, Hermann Weyl, inspired by this geometric formalism, attempted to unify gravity and electromagnetism by positing a local scale or "gauge" invariance~\cite{Weyl:1918ib}. While this specific geometric interpretation failed, the conceptual kernel was revolutionary. After Fritz London identified the crucial link to phase transformations in quantum mechanics~\cite{London:1927fk}, Weyl, in 1929, repurposed his idea~\cite{Weyl:1929fm}. He recognized that the invariance of a quantum theory under local phase transformations of the wave function, $\psi \to  e^{i\theta(x)}\psi$, was the true symmetry at play. To uphold this local $U(1)$ symmetry, the electromagnetic potential $A_\mu$ had to be introduced, transforming in a specific way to cancel the offending terms. Thus, the modern concept of gauge symmetry was born, with electromagnetism emerging as its first and simplest exemplar: a force dictated by the demand for local invariance.

Parallel to these developments in "internal" symmetries, nuclear physics demanded new organizational principles. In 1932, Werner Heisenberg, following the discovery of the neutron, proposed that the proton and neutron were two states of a single "nucleon," differentiated by a new quantum number called "isospin."~\cite{Heisenberg:1932dw} Mathematically, this was a global $SU(2)$ symmetry. This successful classification hinted at a deeper unity, but as a global symmetry, it lacked a dynamical mechanism.

The great leap forward came in 1954, when Chen Ning Yang and Robert Mills took Heisenberg's global $SU(2)$ isospin symmetry and boldly demanded it be made local~\cite{Yang:1954ek}. Following Weyl's template, they constructed a theory where invariance under local $SU(2)$ transformations required the existence of a triplet of massless gauge bosons. This was the birth of non-Abelian Yang-Mills theory. While its immediate application to isospin was not successful, the mathematical framework was of breathtaking power and generality, providing the blueprint for the fundamental forces.

The subsequent explosion of "strange" particles led Murray Gell-Mann and Yuval Ne'eman to introduce the Eightfold Way~\cite{Gell-Mann:1964ook}, a classification scheme based on the $SU(3)$ flavor symmetry. This global symmetry brilliantly predicted the $\Omega^-$ baryon. Gell-Mann and George Zweig then proposed that all hadrons were composed of three fundamental constituents—quarks—transforming under $SU(3)$. However, deeper inconsistencies, such as the spin-statistics problem of the $\Delta^{++}$ baryon, forced the introduction of a new, hidden quantum number: color. It was realized that this "color" charge was the true basis for the strong force, governed by an exact $SU(3)$ gauge symmetry. This theory, Quantum Chromodynamics (QCD)~\cite{Politzer:1973fx,Gross:1973id,Fritzsch:1973pi}, is a direct application of the Yang-Mills principle, with the eight gluons serving as its self-interacting force carriers.

Simultaneously, the weak interaction was undergoing its own transformation. The analogy between beta-decay and electromagnetism suggested a gauge structure. Building on the concept of weak isospin for the left-handed electron and neutrino doublet, Sheldon Glashow in 1961 proposed a unified model of the electromagnetic and weak forces based on the gauge group $SU(2)_L \times U(1)_Y$~\cite{Glashow:1961tr}. This model elegantly grouped the force carriers, but a critical problem remained: the local gauge symmetry demanded massless force carriers, contradicting the short range of the weak force.

The final, triumphant piece was provided in 1967 by Steven Weinberg and Abdus Salam~\cite{Weinberg:1967tq,Salam:1964ry}, who incorporated the Higgs mechanism~\cite{Higgs:1964ia,Higgs:1964pj,Guralnik:1964eu} into Glashow's model. In this framework, the $SU(2)_L \times U(1)_Y$ gauge symmetry is exact in the underlying Lagrangian, but the vacuum state is not symmetric. A scalar Higgs field acquires a nonzero vacuum expectation value, spontaneously breaking the symmetry down to the electromagnetic $U(1)_{\textrm{EM}}$. In this process, three of the four gauge bosons $(W^+, W^-, Z^0)$ acquire mass, while the photon remains massless. This electroweak theory, confirmed by the discovery of the predicted bosons, became the cornerstone of the Standard Model.

This rich historical narrative underscores a single, powerful theme: the evolution of particle physics has been the evolution of symmetry, from a passive tool for classification to an active dynamical principle. The language for articulating this principle is group theory, and its most profound manifestation is the gauge principle. This note will now delve into the formal structure of this magnificent edifice, first by exploring the mathematical lexicon of group theory, then by detailing the engine of the gauge principle, and finally by synthesizing these concepts into the unified framework of the Standard Model.

\section{Group theory}\label{gauge}

In this section, we will introduce the general concepts in group theory and representation theory, and discuss three common groups used in particle physics -- the symmetric groups, the $SU(N)$ group, and the Poincaré group. Besides, the charge conjugate group is also discussed.

\subsection{Basic concept}
The group is a set $G$ equipped with a binary operation $\cdot$, usually called group multiplication, that maps two elements into another element, such that it satisfies the following four properties:
\begin{itemize}
\item Closure: $\forall\ a,b\in G$, $a\cdot b\in G$,
\item Associativity: $(a\cdot b)\cdot c = a\cdot (b\cdot c)$,
\item Identity element: $\exists\ e\in G$, such that $e\cdot a =a ,\ \forall\ a\in G$,
\item Inverse element: $\forall\ a\in G$, $\exists\ a^{-1}\in G$ such that $a\cdot a^{-1}=a^{-1}\cdot a = e$.
\end{itemize}
A group that contains finite elements is called a finite group, and the number of elements in the group is called the \textbf{order} of the group. We call a group \textbf{Abelian} if the group multiplication is commutative for each pair of elements. 
One of the simplest finite group is the $Z_2$ group, which contains two elements $\{-1,1\}$, and the binary operation is the ordinary multiplication for integers. One can verify that all four properties above are satisfied, and it is an Abelian group. In particle physics, the $Z_2$ group can be used to describe the parity and charge conjugate transformations, and is also used to forbid certain types of interactions to stabilize the dark matter.

The most important class of finite groups is the symmetric group. The symmetric group $S_n$ contains the elements of permuting $n$ objects. Its element $p$ can be written as a two-row array of the form:
\begin{equation}
    p = \begin{pmatrix}
    1 & 2 & \dots & n\\
    p_1 & p_2 &\dots & p_n
    \end{pmatrix},\label{eq:perm}
\end{equation}
which specifies the moving of the $i$-th object to the $p_i$-th slot. It is obvious that the order the the column is not important in this notation, what is essential is the concrete map of $p_i$. 
To be more specific, one element of $S_5$ can be written as:
\begin{eqnarray}
   \begin{pmatrix}
    1 & 2 & 3 & 4 & 5\\
    2 & 3 &1 & 5 & 4
    \end{pmatrix},
\end{eqnarray}
where the 5 elements are permuted in two cycles: $1\to 2\to 3\to 1$ and $4\to 5\to 4$, so in this way, we can abbreviate the above elements in the cycle notation: $(123)(45)$. To recover the two-row notation, one can write the first row from 1 to n in order, and then read off the number next to $i$ in a cycle and plug it into the second row in the $i$-th column.  
One advantage of the two row notation is that it manifests the multiplication rule as follows:
\begin{eqnarray}
\text{for }r=\begin{pmatrix}
    1 & 2 & \dots & n\\
    r_1 & r_2 &\dots & r_n
    \end{pmatrix},\ \text{then }r\cdot p =\begin{pmatrix}
    1 & 2 & \dots & n\\
    r_{p_1} & r_{p_2} &\dots & r_{p_n}
    \end{pmatrix}.
\end{eqnarray}
The importance of the symmetric group is obvious for particle physics, as the wave function or amplitude must be totally anti-symmetric (symmetric) under the permutation of the identical fermionic (bosonic) particles. It is also very useful for organizing the flavor structure of the higher-dimensional operators for different effective field theories.

A group can also contain infinite elements. For example, the integer set with the plus operation forms a group, which contains countably infinite elements. 
The set of $n \times n$ unitary matrices with determinant equal to one forms a group under matrix multiplication, known as the $SU(n)$ group. The elements of this group can be specified by a set of continuously varying real parameters. In general, matrix multiplication within $SU(n)$ is non-commutative, making it a \textit{non-Abelian} group for all $n > 1$. These $SU(n)$ groups play a central role in describing particle interactions in quantum field theory, as will be explored in detail in the next chapter.

\subsection{Equivalence relation}
Mathematicians often employ the divide-and-conquer approach to break down a complex problem into smaller and simpler components. The same idea can be applied to the study of group structures. In the following, we will discuss several ways to decompose group elements into more fundamental parts.

One way to partition the group elements is by finding the \textbf{conjugate class}. A group element $g$ is said to be conjugate to $h$ if one can find a group element $p$ such that $g=php^{-1}$. The conjugate defines an equivalence relation among the group elements, as it is straightforward to verify that this relation satisfies reflexivity, symmetry, and transitivity. A set of group elements that are conjugate to each other forms a conjugate class of the group. 

Another important concept is \textbf{subgroup}, which is a subset $H$ of the original group $G$ that also forms a group. One can verify that given a subgroup $H$, then $aHa^{-1}\equiv \{a h a^{-1}, h\in H\}$ is also a group, called the conjugate subgroup of $H$. If the conjugate subgroups of a subgroup are all the same for all elements in the original group $G$, then the subgroup $H$ is called \textbf{invariant subgroup} (\textbf{or normal subgroup}). The original group and the identity element subgroup $\{e\}$ are the two trivial subgroups of $G$. We call a group \textbf{simple} if it does not have a non-trivial invariant subgroup, and \textbf{semi-simple} if it does not have a non-trivial Abelian invariant subgroup. 

With the concept of subgroups, one can divide the group elements into different \textbf{cosets}. A left (right) coset is a set that is formed by left (right) multiplying the group elements $g\in G$ to a subgroup $H$, denoted as $gH\equiv \{gh_i | h_i \in H\}$.
The coset itself need not be a subgroup, as it may not have an identity element. However, one can prove that different group elements will belong to distinct cosets, which means two cosets will be either identical or have no common element at all. 
Furthermore, if the subgroup $H$ is an invariant subgroup, one can show that the set of cosets endowed with the multiplication rule $g_1 H \cdot g_2 H \equiv (g_1\cdot g_2) H$ forms a group. We call this group the \textbf{quotient group}, denoted by $G/H$.
The name ``quotient", exactly like the quotient or modulus operation in the set of integers, gives you the way to define a equivalence relation of two elements in the original group according to whether they belong to the same coset.

On the other hand, one can also define the \textbf{direct product} of two groups $G_1\times G_2$, of which the elements are formed by the ordered pair of the form $(g_1, g_2)$ for $g_1\in G_1$ and $g_2\in G_2$, and the multiplication rule is defined by: $(g_1,g_2)\cdot (g_1',g_2')=(g_1\cdot g_1', g_2\cdot g_2')$.
In addition to the direct product, a more general way to compose a new group with two old ones is \textbf{semi-direct product}. 
A group $G=A\rtimes_\phi B$ is a semi-direct product of an invariant subgroup $A$ and a subgroup $B$, with a $B$-element dependent automorphism $\phi$ of $A$\footnote{Automorphism is an isomorphism that maps a group $G$ to itself, which the set of different maps form a group called Automorphism groups, denoted as ${\rm Aut}(G)$. 
The group element $g\in G$ with conjugate operation labels a subset of automorphism $\phi_g(g')=gg'g^{-1}$, which is called the inner automorphism ${\rm Inn}(G)=G$. 
The outer automorphism ${\rm Out}(G)={\rm Aut}(G)/{\rm Inn}(G)$ describes the true intrinsic symmetry among the group elements in $G$ that cannot be captured by the trivial reshuffling by the conjugate operation.} . 
Similar to the direct product, each element can be uniquely represented by the pair $(a,b)$, while the product rule is twisted by the $\phi$: $(a_1,b_1)\cdot (a_2,b_2)\equiv (a_1\phi_{b_1}(a_2), b_1b_2)$. In the case where $\phi_b$ is an inner automorphism, $\phi_b(a)=bab^{-1}$.
On the application of high energy physics, the Pioncar\'e group is a semi-direct product of the translation group and the Lorentz group in Minkowski space, and the irreducible unitary representation of which is first analyzed by Wigner using the induced representation method~\cite{Wigner:1939cj} and later generalized to the local compact Lie group by Mackey~\cite{Mackey1970}.

Up to now, we have studied how to partition the group according to different two equivalence relations.  To further study the structure of groups and identify the similarity between different groups, we introduce the concept of 
\textbf{homomorphism}. A homomorphism from one group $G$ to another $G'$ is a map $f$ from $G$ to $G'$, such that it preserves the multiplication rule of the group:
\begin{equation}
   f: G\to G', s.t.\  f(g_1\cdot g_2) = f(g_1)\cdot f(g_2), \ \forall\ g_1,g_2\in G.
\end{equation}
Pay attention that the $\cdot$ on the left-hand side and the one on the right-hand side are multiplications in $G$ and $G'$, respectively. 
If there is a homomorphism $h$ between two groups $G$ and $G'$ that is also bijective, then we call the two groups \textbf{isomorphic} to each other, and the map is called an isomorphism, which is a special case of homomorphism. The two groups are considered to be identical, denoted by $\cong$, if they are isomorphic to each other, and isomorphism is considered to be the highest level of equivalence at the level of group.

One of the most important theorems based on homomorphism is the so-called \textbf{isomorphism theorem}. Given a homomorphism $f: G\to G'$, we call the image of $f$ as ${\rm Im}f$, which is a subset of $G'$, and the kernel of $f$, ${\rm ker}f$, defined as the subset of elements in $G$ that are mapped to identity elements in $G'$. It can be easily shown that ${\rm ker}f$ is an invariant subgroup of $G$, and the isomorphism theorem tells that ${\rm Im}f \cong G/{\rm ker}f$. 
Interestingly, the definition of cosets and their multiplication rule in a quotient group automatically defines a homomorphism between the original group $G$ and qoutient group $G/H$.
Since $H=e H$ is the identity element in the quotient group, $H$ is the kernel of this coset map.

\subsection{Lie group and Lie algebra}
Lie groups are used to describe continuous transformations on a physical system, where the group elements are parameterized with continuous variables and they form smooth manifolds.
The most common Lie groups in high-energy physics are the matrix Lie groups.  
For example, the rotation in the $R^N$ forms the group $SO(N)$, and the gauge groups that describe the symmetries in the Standard Model are $SU(N)$ groups -- the $n\times n$ unitary matrices with determinant equal to one.

Depending on the topological properties of the group manifold, the Lie group can be classified as compact or non-compact. Heuristically, the ``compactness" here means that you can ``warp around" the group and no direction in the group manifold goes off to infinity. For example, the group manifold of $SU(2)$ is $S^3$, therefore compact, while that for the group $SL(2,\mathbbm{C})$, consisting of $2\times 2$ complex matrices with unity determinant, is $R^3\times S^3$, which is non-compact.
The compactness of a Lie group determines the dimensionality of its unitary irreducible representations. For compact Lie groups, these representations are finite-dimensional, whereas for non-compact groups, they are infinite-dimensional. Moreover, the methods used to construct the unitary irreducible representations in the two cases differ significantly.

As the matrix Lie group can be viewed as a smooth manifold, one can largely study its properties by analyzing the infinitesimal transformation near the identity. 
Because each group element can be linked to a sequence of infinitesimal transformations from the identity by an exponential map.  
This set of infinitesimal transformations, from the geometric point of view, is the vector in the tangent space at the identity element $T_e G$. 
If locally the group element can be parameterized with a set of real parameters $\boldsymbol{\alpha}$ as $g(\boldsymbol{\alpha})$, with $g(\boldsymbol{0})=e$, then we can define the generator of the Lie group:
\begin{eqnarray}
\label{eq:generator}
T_i = -i\left.\frac{\partial g}{\partial\alpha_i}\right|_{\boldsymbol\alpha=\boldsymbol{0}},
\end{eqnarray} 
they form a basis of the tangent space at $e$.
Equipped with the commutation relation:
\begin{eqnarray}
[T_i, T_j] = if^{ijk}T_k,
\end{eqnarray}
the tangent space is promoted to the Lie algebra $\mathfrak{g}$ of the Lie group $G$, where $f^{ijk}$ is called the structure constant, up to a change of basis, uniquely fixes a Lie algebra. With this commutation relation, one can recover the multiplication rule for finite transformation with Baker–Campbell–Hausdorff formula:
\begin{eqnarray}
e^X e^Y=e^{X+Y+\frac{1}{2}[X,Y]+\frac{1}{12}\left([X,[X,Y]]+[Y,[Y,X]]\right)}+\dots
\end{eqnarray}
Both the commutator of generators and the structure constant satisfy the \textbf{Jacobian identity}\footnote{Here we start with matrix Lie groups, and the generators can be viewed as matrices, and the commutator $[A,B]$ can be defined with the matrix multiplication $AB-BA$, so the Jacobian identity can be derived by expanding the commutator. However, when starting with Lie algebra, where the multiplication of two generators is not defined, then the Jacobian identity should be viewed as a part of the definition of the Lie bracket.}:
\begin{eqnarray}
&&[T_i,[T_j, T_k]]+[T_k,[T_i, T_j]]+[T_j,[T_k, T_i]]=0,\\
&&f_{bcd}f_{ade}+f_{abd}f_{cde}+f_{cad}f_{bde}=0.
\end{eqnarray}
The study of Lie algebras provides a powerful way to understand the corresponding Lie groups, including the derivation of irreducible representations and the classification of different Lie groups.
For example, one can define \textbf{Casimir operators} that commute with all the generators, and due to the Schur lemma, these operators must be proportional to the identity operator for an irreducible representation.
Therefore, the set values of Casimir operators can be used to label different irreducible representations of the group.
In this review, we will not go further into the heavy machinery developed by Cartan, and refer the interested reader to conventional textbooks~\cite{Georgi:2000vve,Hall:2015xtd,Zee:2016fuk}. 

Even though Lie algebras can provide local information about the Lie group, they fail to capture the global structure of the Lie group. Every connected Lie group $G$ with a given Lie algebra $\mathfrak{g}$ is a quotient of a unique simply connected Lie group $\tilde{G}$ having the same Lie algebra, that is,
\begin{equation}
    G=\tilde{G}/\Gamma,
\end{equation}
where $\Gamma$ is the discrete center subgroup\footnote{A center subgroup is a group that contains elements that are commutative with any elements in the group. }.
For example, $SU(2)$ and $SO(3)$ have the same Lie algebra, with the same structure constant $f^{ijk} = \epsilon^{ijk}$, where $\epsilon^{ijk}$ is the Levi-Civita tensor.
For a rotation in 3-D, the rotation angle $\theta$ and $\theta+2\pi$ give the same group element, while the same $2\pi$ ``rotation'' gives an extra minus sign in $SU(2)$. Therefore, we say, $SU(2)$ is a double-covering group of $SO(3)$, also denoted as $SO(3)\cong SU(2)/Z_2$.
From the representation point of view, this tells us that the $SO(3)$ cannot have half-integer representations of the corresponding Lie algebra $\mathfrak{so}(3)=\mathfrak{su}(2)$.
The same argument applies to the Standard Model gauge group, as we classify the matter fields according to the representation of the Lie algebra.
The representation of matter fields allows four distinct global structures of the gauge groups characterized by $\tilde{G} = SU(3)\times SU(2)\times U(1)$ and four different $\Gamma$:
\begin{eqnarray}
    \Gamma_{1,2,3,4}= Z_6, Z_3, Z_2, \text{ and}\ \{I\}.
\end{eqnarray}
This was first pointed out in the textbook Ref.~\cite{ORaifeartaigh:1986agb}, and recently reviewed by Ref.~\cite{Tong:2017oea}, where the author pointed out the relation between the center subgroup and the 1-form symmetry that acts on the line operators, which results in a different pattern of charge quantization for the particles and magnetic monopoles.
See Ref.~\cite{Agrawal:2017cmd,Koren:2022axd,Wang:2022eag,Cordova:2022fhg,Davighi:2023iks,Reece:2023iqn,Choi:2023pdp,Cordova:2023her,Alonso:2024pmq,Khoze:2024hlb,Li:2024nuo,Koren:2024xof,Gould:2024zed,Delgado:2024pcv,Davighi:2025iyk,Alonso:2025rkk} for more discussions on the phenomenological consequences and application of different global structures of the Standard Model gauge groups and beyond. 

\subsection{Representation theory}
In high-energy physics, representation theory is essential because it provides the bridge between abstract symmetry groups and the physical particles and fields that we observe. While group theory identifies the underlying symmetry transformations of a physical system—such as the gauge symmetries $SU(3)$, $SU(2)$, and $U(1)$ in the Standard Model—representation theory tells us how particles, as the building blocks of the theory, transform under these symmetries. 
Each type of particle corresponds to a specific representation of the symmetry group: for instance, quarks transform under the fundamental representation of $SU(3)$, while gluons belong to its adjoint representation. 
These transformation properties determine a particle’s quantum numbers, such as charge, color, and isospin, and dictate how different particles can interact or combine. 
To construct a theory that satisfies certain symmetry amounts to constructing a singlet representation with those building blocks.
In this sense, representation theory is a fundamental tool for uncovering the concrete form of the theory under certain symmetry principle.

A \textbf{representation} of a group $G$ is a homomorphism of the group elements to the invertible linear operator $U(g)$ that acts on the vector space $V$. Once a basis $\{\boldsymbol{e}_i\}$ in the $V$ is determined, $U(g)$ can be realized as a concrete representation matrix (rep-mat) $D(g)^j{}_i$, such that:
\begin{equation}
    U(g)\boldsymbol{e}_i = \sum_j \boldsymbol{e}_j D(g)^j{}_i.
\end{equation}
Sometimes physicists also call the matrix $D(g)$ or even the vector space $V$ as the representation of the group.
Given this definition of a representaiton, we  define the dimension of the representation as the dimension of the vector space.

The two representations $U(g)$ and $U'(g)$ are \textbf{equivalent} if there exists an invertible transformation $S$, such that $U'(g)=S U(g)S^{-1}$ for all $g\in G$.
This merely corresponds to a basis transformation in $V$.
The representation is \textbf{irreducible} if it does not contain any invariant subspace $W$ such that $\forall g\in G,\ U(g)w \in W,\ \forall w\in W$.
A representation is \textbf{unitary} if $U(g)^\dagger= U(g)^{-1}$, it is particularly important to physicists as it preserves the probability interpretation of a physical state under the symmetry transformation. 
From two known representations $U_1(g)$ and $U_2(g)$  and the corresponding vector space they act on  $V_1$ and $V_2$, the \textbf{direct sum} representation can be construced with $U(g)\equiv U_1(g)\oplus U_2(g)$ that acts on the new vector space $V=V_1\oplus V_2$, and the rep-mat can be put into block diagonal form 
\begin{equation}
    D(g)=\begin{pmatrix}
    D_1(g)& \boldsymbol{0}_{n_1\times n_2}\\
    \boldsymbol{0}_{n_2\times n_1} & D_2(g)
    \end{pmatrix},
\end{equation}
where $D_{1,2}(g)$ are rep-mat of $U_{1,2}$, and $n_{1,2}$ are dimension of $V_{1,2}$.
We state without proof that any representation of finite groups and compact Lie groups can be decomposed as a direct sum of irreducible representations.
Apart from the direct sum, one can also construct the new representation with \textbf{direct product}. The new vector space is formed by $V=V_1\otimes V_2$, with basis vector $\{\boldsymbol{e}^1_i\otimes \boldsymbol{e}^2_j\}$, and the action of $U(g)=U_1(g)\otimes U_2(g)$ on the basis is defined to be:
\begin{eqnarray}
U(g)(\boldsymbol{e}^1_i\otimes \boldsymbol{e}^2_j) \equiv (U_1(g)\boldsymbol{e}^1_i)\otimes (U_2(g)\boldsymbol{e}^2_j)=\boldsymbol{e}^1_k\otimes \boldsymbol{e}^2_l D^1(g)_{ki}D^2(g)_{lj}.
\end{eqnarray}

To tell different representations apart, we define the \textbf{character} of a representation $\chi_\mu(g)={\rm Tr}D_\mu(g)$, where $\mu$ labels the representation, and $g$ labels the group element. However, due to the nature of the trace, the character is a function of conjugate classes only.
Therefore, the character can also be viewed as a vector with conjugacy classes as indices. It can be proved that the characters satisfy the following orthonormal conditions for finite groups:
\begin{eqnarray}
\frac{n_i}{n_\mu}\sum_i \chi^\dagger_{\mu }(i)\chi_{\nu }(i) = \delta_{\mu\nu},
\end{eqnarray}
where $n_i$ and $n_\mu$ are the number of elements in the conjugacy class $i$ and the dimension of the representation $\mu$ respectively.
This formula is powerful for computing the multiplicity of certain irreducible representation $a_\mu$ in a given representation $\mu$, if the rep-mat and thus the character $\chi(i)$ are known:
\begin{eqnarray}
a_\mu = \frac{n_i}{n_\mu}\sum_i \chi^\dagger_{\mu }(i)\chi(i)
\end{eqnarray}
This formula can be generalized to a compact Lie group by replacing the summation with an integral with proper normalization.
For example, for SU(2), the rotations with the same angle $\theta\in (0,\pi)$ but around different axis consist of a conjugacy class, with character $\chi_j(\theta)$, where $j$ a half-integer labeling the irreducible representation:
\begin{eqnarray}
\chi_j(\theta) = \frac{\sin((2j+1)\theta)}{\sin\theta}.
\end{eqnarray} 
The integration measure on the classes yields $\mu(\theta) = 2\sin^2\theta/\pi$, and the orthonormal condition is given by:
\begin{eqnarray}
\int_0^\pi\chi^\dagger_j(\theta)\chi_k(\theta)  \mu(\theta)d\theta = \delta_{jk}.
\end{eqnarray}
This approach forms the foundation in counting the number of independent operators of effective field theory using the Hilbert series method~\cite{Feng:2007ur,Hanany:2010vu,
Lehman:2015via,Lehman:2015coa,Henning:2015daa,Henning:2015alf,Graf:2020yxt,Sun:2022aag,Graf:2022rco}.

\subsubsection{Symmetric group}
We define \textbf{group algebra} $\tilde{S}_n$ as the vector space consisting of complex linear combinations of group elements of $S_n$, $\tilde{S}_n=\{r=\sum_i r_ig_i\ r_i\in \mathbbm{C},\  g_i \in S_n\}$. where we have chosen the group elements $g_i$ as the basis vectors. 
Since it is an algebra, we have the multiplication of the two elements $r,p$ in addition to the addition, which can be defined naturally using the multiplication rule of group elements
:
\begin{eqnarray}
   r\cdot p = \sum_i r_i g_i\cdot \sum_j p_jg_j = \sum_{i,j,k }r_ip_j\Delta^k_{ij}g_k.\label{eq:regrep}
\end{eqnarray}
Given that $g_i g_j=g_m$, then $\Delta^k_{ij}$ is equal to zero unless $k=m$. $\Delta^k_{ij}=(\Delta_i)_{kj}$ can be regarded as the rep-mat of the group element $g_i$,as one can verify $(\Delta_i)_{pq}(\Delta_j)_{qr}=(\Delta_m)_{pr}$ by acting $g_ig_j$ and $g_m$ one the basis vector.
This representation, defined on the group algebra space, is called the \textbf{regular representation}, it is not only decomposable but also contains all the irreducible representations for the $S_n$ group.
From eq.~\eqref{eq:regrep}, one can observe the dual rule of the elements of the group algebra, as the operator that acts on the vector space and as the vector itself.

To study the irreducible representation of $S_n$, we define the partition and Young diagram as follows.
A \textbf{partition} of a integer $n$ is a set of integer $\lambda=[\lambda_{i=1,\dots k}]$ such that $\sum_i\lambda_i=n$ and $\lambda_i\geq \lambda_{i+1}$.
We can draw a stacked boxes with numbers of boxes in each row equal to $\lambda_i$ from top to bottom, which is called the \textbf{Young diagram} of the corresponding partition.
For example, the partition of integer $5$ -- $[3,2]$ and $[4,1]$ correspond to two Young diagrams:
\begin{eqnarray}
    \yng(3,2)\quad  \yng(4,1).
\end{eqnarray}
This kind of Young diagram with the number of boxes not decreasing from top to bottom and from left to right is called \textbf{normal Young diagram}. 
When filling the Young diagram with numbers, one obtains the \textbf{Young tableau}, and a \textbf{standard Young tableau} is the one with numbers in each row and each column in increasing order.
For example, followings are the four standard Young tableaux of the Young diagram of the shape $[4,1]$
\begin{eqnarray}
    \young(1234,5)\quad \young(1235,4)\quad \young(1245,3)\quad \young(1345,2).
\end{eqnarray}
We state without proof here that the irreducible representations of the $S_n$ group are characterized by the Young diagram of different shapes, and the dimension of the representation is equal to the number of Standard Young tableaux when filling the diagram with numbers $1$ to $n$ without repetition. As illustrated above, the dimension of the $[4,1]$ irreducible representation of $S_5$ is $4$.
In general, we have the formula for the dimension of the Young diagram $\lambda$ of the $S_n$ group:
\begin{eqnarray}
    d_{\lambda}=\frac{n!}{\Pi_{ij}h_{ij}},\label{eq:dsym}
\end{eqnarray}
where $h_{ij}$ is the so-called hook number of the box ast $i$-th row and $j$-th column, which is defined to by adding the number of remaining boxes to the right and to the bottom plus 1.
For the Young diagram $[4,1]$, we can fill the hook numbers in each box to obtain the hook number tableau:
\begin{eqnarray}
    \young(5321,1).
\end{eqnarray}

To obtain the rep-mat of an irreducible representation, we need to find the corresponding irreducible subspace in the group algebra.
This leads us to the definition of the Young symmetrizer of a Young tableau:
\begin{eqnarray}
    {\cal Y}={\cal P}{\cal Q},
\end{eqnarray}
where ${\cal P}({\cal Q})$ is the product of total symmetrization $P_i$ (anti-symmetrization $Q_j$) of each row (column):
\begin{eqnarray}
    {\cal P}=\Pi_i P_i,\quad {\cal Q}=\Pi_j Q_j.
\end{eqnarray}
For example, for the Young tableau:
\begin{eqnarray}
    \young(123,45),
\end{eqnarray}
we have:
\begin{eqnarray}
    &&P_1 = E+(12)+(13)+(23)+(123)+(132),\quad
    P_2 = E+(45),\\
    &&Q_1= E-(14),\quad Q_2 = E-(25),\quad Q_3=E.
\end{eqnarray}
Obviously, the Young symmetrizer of $\cal Y$ is an element of the group algebra, and rescaling it with a factor of $\hat{{\cal Y}}=(d_\lambda/n!){\cal Y}$, one can prove that $\hat{{\cal Y}}\hat{{\cal Y}}=\hat{{\cal Y}}$, which is an primitive idempotent of the group algebra.
One of the important properties of the primitive idempotent is that it can generate the subspace of the irreducible representation by left acting on the group element in $S_n$:
\begin{eqnarray}
   V_{\lambda} = {\rm span}\{\ g_i\hat{{\cal Y}}_{\lambda}\ |\  g_i\in S_n\}.
\end{eqnarray}
For example, for the $S_3$ group, the $[2,1]$ irreducible representation of dimension 2 is generated by the Young symmetrizer:
\begin{eqnarray}
   \hat{\cal Y}_{\tiny \young(12,3)}\equiv b^{[2,1]}_1=\lbrace E + (12)-(13)-(132)\rbrace/3,\label{eq:b1}
\end{eqnarray}
and the other basis vector is obtained by acting the group element $(23)$ on $\hat{{\cal Y}}$:
\begin{eqnarray}
    b^{[2,1]}_2=(23)\hat{{\cal Y}}=\lbrace (23)+(132)-(123)-(12)\rbrace/3.\label{eq:b2}
\end{eqnarray}
Acting with other group elements results in the group algebra elements that can be expressed as a linear combination of these two basis vectors.
The rep-mat of this irreducible representation can be obtained directly by reading out the coefficient in front of these two basis vectors after acting by the group elements.
For the example, with explicit expression for $b^{[2,1]}_{1,2}$ in eq.~\eqref{eq:b1} and \eqref{eq:b2}, the rep-mat of the generator of the $S_3$ is given by:
\begin{eqnarray}
    D^{[2,1]}[(12)]=\begin{pmatrix}
        1 & -1 \\
        0 & -1
    \end{pmatrix},\quad 
    D^{[2,1]}[(123)]=\begin{pmatrix}
        -1 & 1 \\
        -1 & 0
    \end{pmatrix}.
\end{eqnarray}
In this way, we can find all the rep-mat of any irreducible representation of the $S_n$ group.
It can be shown that the regular representation defined on the group algebra contains all the irreducible representations of the symmetric group, and the multiplicity of each irreducible representation is equal to the dimension of that representation.
Readers are encouraged to refer to textbooks~\cite{Tung:1985iqd,ma2007group} for more details.

\subsubsection{$SU(N)$ group}
We use the tensor method to construct the irreducible representation of the $SU(N)$ group, where each representation is obtained by the tensor product of the fundamental representation.
The \textbf{fundamental representation} of the $SU(N)$ group is defined on a $n$-dimensional complex vector space, and the matrix $g_{ij}\in SU(N)$ is mapped to itself. For any vector $\boldsymbol{v}$ with component $v_i$, the resulting vector $\boldsymbol{v}'=U_{\rm fund}(g)\boldsymbol{v}$ has components:
\begin{eqnarray}
v'_i = g_{ij}v_j.
\end{eqnarray} 

A generic tensor constructed with the direct product of $n$ fundamental representations is usually reducible.  
The irreducible subspace can be obtained by restricting tensor indices to have definite permutation symmetries. 
This is rooted in the fact that the permutation on the indices of the tensor and the $SU(N)$ transformation are commutative. 
Based on this observation, the simplest irreducible representation can be obtained by totally symmetrizing or the totally antisymmetrizing the tensors indices. In terms of Young diagrams, they are represented by a one-row and a one-column diagram with $n$-boxes, each box represents an index of the fundamental representation.
For more general cases, a Young diagram of mixed symmetry corresponds to the subspace obtained by acting the corresponding Young symmetrizer on the indices.
For example, for a generic tensor of three fundamental indices $\Theta_{a_1a_2a_3}$, one of the invariant subspace of mixed symmetry $[2,1]$ can be obtained by exerting the Young symmetrizer on the tensor:
\begin{eqnarray}
\Theta_{a_1a_2a_3}^{\tiny\young(12,3)}={\cal Y}[{\tiny\young(12,3)}]\Theta_{a_1a_2a_3}=\Theta_{a_1a_2a_3}+\Theta_{a_2a_1a_3}-\Theta_{a_3a_2a_1}-\Theta_{a_3a_1a_2}.\label{eq:ytab}
\end{eqnarray}
where we have defined the following rule for a permutation $p$ in eq.~\eqref{eq:perm} acting on the tensor indices:
\begin{eqnarray}
    p\ \Theta_{a_1a_2\dots a_n} = \Theta_{a_{p_1}a_{p_2}\dots a_{p_n}}.
\end{eqnarray}
As the number of indices of a tensor is unlimited, one can have a Young diagram of any number of boxes, while the only restriction is that the number of columns cannot exceed $N$. Because each column represents a total anti-symmetrization of the indices in that column, while indices can take at most $N$ different number, when two indices takes the same number, the anti-symmetrization render the expression equal to zero. 
In this sense, any column with $N$ box can be stripped off, and identify the Young diagrams before and after the stripping. For example following Young diagrams all represent the fundamental representations for the $SU(2)$ group:
\begin{eqnarray}
\yng(1),\quad \yng(2,1),\quad \yng(3,2)    
\end{eqnarray}
The basis vectors of the irreducible representation are given by the semi-standard Young tableau of the corresponding Young diagram by filling the number from $1$ to $N$. 
Semi-standard Young tablueax are those with numbers not decreasing from left to right and strictly increasing form top to bottom. For example, for $[2,1]$ representation of $SU(3)$, the basis vector is obtained by:
\begin{eqnarray}
    \young(11,2),\ \young(11,3), \ \young(12,2),\ \young(12,3)\  \young(13,2),\ \young(13,3),\ \young(22,3), \ \young(23,3).\label{eq:ytab2}
\end{eqnarray}
Reader should not confuse the semi-standard Young tableaux here with the standard Young tableau in eq.~\eqref{eq:ytab}. Here, the number in the Young tableaux represent the values that the indices of $a_{1-3}$ take.
For instance, the first Young tableau in eq.~\eqref{eq:ytab2} corresponds to the specific tensor component with $a_1=1$, $a_2=1$, and $a_3=2$ in eq.~\eqref{eq:ytab}.
The dimension of the representation can also be read out from the Young diagram with the following formula:
\begin{eqnarray}
    d_{[\lambda]}(SU(N)) = \prod_{ij}\frac{N+j-i}{h_{ij}},
\end{eqnarray}
where $h_{ij}$ are hook number defined below eq.~\eqref{eq:dsym}, again $i$ and $j$ indicate the $i$-th row and $j$-th column of the Young diagram.

A common problem in the application of particle physics is to count the multiplicities of different irreducible representations of tensor product representations.  
The \textbf{Littlewood-Richardson rule} \cite{littlewood1934group,OKADA1998337} gives a diagrammatic solution to the problem.
For the tensor product of two irreducible representations $[\lambda]$ and $[\mu]$, take the one with less box e.g. $[\mu]$, and fill the number $i$ to the i-th row from top to bottom, then paste the box together with the number of each row from $[\mu]$ to the right of the diagram $[\lambda]$. Each time finishing one row of boxes, check the following 4 conditions:
\begin{enumerate}
    \item The resulting Young diagram is still a normal Young diagram.
    \item No boxes with the same number are in the same column.
    \item From the right to left and from top to bottom, counting the number of boxes of each number. At each step, the number of boxes filling $j$ is not larger than the number of boxes filling $k$ for any $j>k$.
    \item The number of boxes in the column is no larger than $N$.
\end{enumerate}
Finally, the number of occurrences of each type of Young diagram is the multiplicity of the corresponding irreducible representation. 
For example, the tensor product of two $[2,1]$ of $SU(3)$ proceeds as follows:
\begin{eqnarray}
   {\tiny \yng(2,1)}\otimes {\tiny \young(11,2)}= {\tiny\young({{}}{{}}11,{{}}2)}\oplus  {\tiny\young({{}}{{}}11,{{}},2)}\oplus {\tiny\young({{}}{{}}1,{{}}12)}\oplus
   {\tiny\young({{}}{{}}1,{{}}1,2)}\oplus
   {\tiny\young({{}}{{}}1,{{}}2,1)}\oplus
   {\tiny\young({{}}{{}},{{}}1,12)}.
\end{eqnarray}

In addition to the Young diagram, one can also use the values of Casimir operators to label the irreducible representation.
The $SU(N)$ group has $N-1$ different Casimir operators, and they can be constructed with the generators and the symmetric and anti-symmetric structure constants.
The most commonly used are 2-nd and 3-rd order Casimir operators:
\begin{eqnarray}
    C_2 = T^aT^a, \quad C_3= d^{abc}T^aT^bT^c,
\end{eqnarray}
where $d^{abc}$ is the symmetric structure constant defined by:
\begin{eqnarray}
     d^{abc}=2{\rm Tr}(\{T_F^a,T_F^b\}T_F^c).
\end{eqnarray}
In general, the $n$-th order Casimir operator is constructed with $n$ generators contracting with a totally symmetric tensor with adjoint indices, and an explicit construction for the $su(N)$ Lie algebra can be found in Ref.~\cite{deAzcarraga:2000ng}.

Another frequently encountered problem in the application of particle physics is to evaluate the trace of generators in computing the scattering amplitude. For the generator of the fundamental representation, $T_F$, we have the following useful identities\footnote{More identities can be found in Ref.~\cite{Haber:2019sgz}}:
\begin{eqnarray}
    {\rm Tr}(T_F^aT_F^b)=\frac{1}{2}\delta^{ab}\mathbbm{1},\quad T_F^aT_F^b=\frac{1}{2}\left[\frac{1}{N}\delta^{ab}\mathbbm{1}+(d^{abc}+if^{abc})T_{Fc}\right], \quad (T_F^a)_{ij} (T_F^a)_{kl} = \frac{1}{2}\left(\delta_{il}\delta_{kj}-\frac{1}{N}\delta_{ij}\delta_{kl}\right).\label{eq:idfund}
\end{eqnarray}
By converting the $d^{abc}$ and $f^{abc}$ into trace of generators,
one can always use eq.~\eqref{eq:idfund} recursively to solve the trace of any number of fundamental generators and adjoint representation since its generator $(T_A^a)_{bc}=-if^{abc}$.
For the anti-symmetric representation $[1,1]$ denoted as a tensor $u_{[ij]}$ with two fundamental indices anti-symmetrized, its generator can be expressed with the generator of the fundamental representation in the following form:
\begin{eqnarray}
    (T^a_{\rm anti})^{i_2 j_2}_{i_1j_1} = \delta_{j_1}^{j_2} (T^a_F)_{ i_1}^{i_2}-\delta_{j_1}^{i_1} (T^a_F)_{i_1}^{j_2}-\delta_{i_1}^{j_2} (T_F^a)_{j_1}^{i_2}+\delta_{i_1}^{i_2} (T_F^a)_{j_1}^{i_2},
\end{eqnarray}
and this form enables one to evaluate its trace with the identities of fundamental representation, while the only modification needed is to divide a factor of 2 for the contraction of each pair of the anti-symmetrized indices.
A similar technique can be applied to the symmetric representation $[2]$, denoted as a tensor $u_{\{ij\}}$. In this case, the generator is written as:
\begin{eqnarray}
    (T^a_{\rm sym})^{i_2 j_2}_{i_1j_1} = (T^{a}_F)^{i_2}_{i_1}\delta^{j_2}_{j_1}+(T^{a}_F)^{j_2}_{j_1}\delta^{i_2}_{i_1}-(T^a_{\rm anti})^{i_2 j_2}_{i_1j_1}.
\end{eqnarray}
For other representations, one can resort to computer code such as Refs.~\cite{Feger:2019tvk,Fonseca:2020vke} for numerical evaluations.


\newcommand{\pg}{Poincar\'e }

\subsection{Poincar\'e Group}

\pg group $\mathcal{G}$ is the maximal group that transfers the reference frames to each other in special relativity \cite{1906RCMP...21..129P}.
Time and space are unified to be the spacetime in special relativity, and their coordinates are of dimension 4. Among the 4 coordinates, one is the time $t$, and the other three are the space coordinates $x,y,z$. Writing the spacetime coordinates by a dimension-4 vector,
\begin{equation}
x^\mu = (t,x,y,z)^T\,,
\end{equation}
called 4-vector or spacetime vector, then the relativity principle requires the quantity $x^2 = x^\mu x_\mu = t^2-(x^2+y^2+z^2)$ is unchanged under the \pg transformations. If we introduce a symmetric tensor
\begin{equation}
	g_{\mu\nu} = \text{diag}(1,-1,-1,-1)\,,
\end{equation}
as the metric of the spacetime, the interval $x^2$ can be regarded as the corresponding inner product,
\begin{equation}
	x^2 = x^\mu x_\mu = x^\mu g_{\mu\nu}x^\nu\,.
\end{equation}
The 4-dimensional spacetime and the metric $g$ form a metric linear space known as Minkowski space, and the metric $g$ is referred to as the Minkowski metric.
Then \pg group $\mathcal{G}$ is the maximal group keepping the Minkowski metric unchanged,
\begin{equation}
g \rightarrow G^TgG = g \,,\quad G\in \mathcal{G}\,.
\end{equation}

The \pg group is a typical application of group theory in particle physics, which covers and broadens most of the topics discussed before, including Lie algebra, semi-direct product, compactness, induced representation, and so on. Besides, the infinite unitary representation of \pg group is closely related to the classification of particles in particle physics~\cite{Wigner:1939cj,Mackey1970}. 


\paragraph{Poincar\'e Algebra} All the transformations in \pg group can be classified into 3 sectors,
\begin{itemize}
	\item 3 rotations along the 3 spatial directions, $x\rightarrow R x$. For example, the rotation along the $z$ axis by angle $\theta$ keeps the $z$ and $t$ coordinates unchanged while rotate the $xy$-plane by $\theta$, whose matrix representative is 
    \begin{equation}
        R_3(\theta) = \left(\begin{array}{cccc}
        1& 0 & 0 & 0 \\
        0& \cos\theta & -\sin\theta & 0 \\
        0& \sin\theta & \cos\theta & 0\\
        0 & 0 & 0 & 1
        \end{array}\right)\,,
    \end{equation}
    where we have written the $\theta$ dependence of the rotation $R_3$ explicitly. The corresponding generator can be obtained by Eq.~\eqref{eq:generator} that
    \begin{equation}
        J_3 = -i\left.\frac{dR_3(\theta)}{d\theta}\right|_{\theta=0} = -i\left( \begin{array}{cccc}
			0 & 0 & 0 & 0 \\
			0 & 0 & 1 & 0 \\
			0 & -1 & 0 & 0 \\
			0 & 0 & 0 & 0 
\end{array}
\right)\,,
    \end{equation}
    Similarly, the generators of the rotations along the $x\,,y$ axis can also be obtained,
    \begin{equation}
	J_1 = -i\left( \begin{array}{cccc}
			0 & 0 & 0 & 0 \\
			0 & 0 & 0 & 0 \\
			0 & 0 & 0 & 1 \\
			0 & 0 & -1 & 0 
\end{array}
\right)\,,\quad 
	J_2 = -i\left( \begin{array}{cccc}
			0 & 0 & 0 & 0 \\
			0 & 0 & 0 & -1 \\
			0 & 0 & 0 & 0 \\
			0 & 1 & 0 & 0 
\end{array}
\right)\,.
\end{equation}
	\item 3 boosts along the 3 spatial directions, $x\rightarrow B x$. Boost transformations are the characteristics that distinguish special relativity from classical mechanics. Opposite to the rotations, a boost along a specific axis alters this coordinate and the time coordinate, while leaving the orthogonal coordinates unchanged. For example, the boost along $z$ axis takes the matrix form that
    \begin{equation}
        B_3(\eta) = \left(\begin{array}{cccc}
            \cosh\eta & 0 & 0 & \sinh\eta \\
            0 & 1 & 0 & 0 \\
            0 & 0 & 1 & 0 \\
            \sinh\eta & 0 & 0 & \cosh\eta
        \end{array}\right)\,,
    \end{equation}
    which depends on the real parameter $\eta$ called rapidity. The hyperbolic functions 
    \begin{equation}
        \cosh\eta = \frac{e^\eta + e^{-\eta}}{2}\,,\quad \sinh\eta = \frac{e^\eta-e^{-\eta}}{2}\,, 
    \end{equation}
    are divergent as $\eta$ becomes large, thus the boost transformations make the \pg group not compact. Similarly, the generators of the boosts along the 3 directions are
    \begin{equation}
	K_1 = -i\left( \begin{array}{cccc}
			0 & 1 & 0 & 0 \\
			1 & 0 & 0 & 0 \\
			0 & 0 & 0 & 0 \\
			0 & 0 & 0 & 0 
\end{array}
\right)\,,\quad 
	K_2 = -i\left( \begin{array}{cccc}
			0 & 0 & 1 & 0 \\
			0 & 0 & 0 & 0 \\
			1 & 0 & 0 & 0 \\
			0 & 0 & 0 & 0 
\end{array}
\right)\,,\quad 
	K_3 = -i\left( \begin{array}{cccc}
			0 & 0 & 0 & 1 \\
			0 & 0 & 0 & 0 \\
			0 & 0 & 0 & 0 \\
			1 & 0 & 0 & 0 
\end{array}
\right)\,,
\end{equation}
	\item 4 translations along the 4 spacetime directions, $x\rightarrow x'=x+b$. If the translation parameter $b$ is infinitesimal, the transformation takes the form
    \begin{equation}
        x'^\mu (b) = x^\mu + \frac{\partial x'^\mu}{\partial x^\nu} b^\nu\,.
    \end{equation}
    The generators $P^\mu$ are defined by the infinitesimal transformation on $x'$,
    \begin{equation}
        P^\mu x'_\nu= -i\frac{\partial x'^\mu (b)}{\partial b^\nu} = -i\partial_\nu x'^\mu\,,
    \end{equation}
    thus they are $P^\mu = -i\partial^\mu$.
\end{itemize}
Therefore, the \pg group is of dimension 10. It is non-compact, and its Lie algebra is spanned by the 10 generators that,
\begin{equation}
    \{J_i\,,\quad K_i\,,\quad P_\mu\}\,.
\end{equation}

To obtain the commutation relations, we consider the rotation and boost generators first. Their commutation relations can be obtained by their matrix representatives,
\begin{align}
	[J_i,J_j] &= i\epsilon_{ijk}J_k\,,\\
	[K_i,K_j] &= -i\epsilon_{ijk}J_k\,,\\
	[J_i,K_j] &= i\epsilon_{ijk}K_k\,.
\end{align}
Or if we express the 6 generators by an antisymmetric tensor $J_{\mu\nu}$, 
\begin{equation}
	J_i = \frac{1}{2}\epsilon_{ijk}J_{jk}\,,\quad K_i = -J_{0i}\,,
\end{equation}
the commutators can be unified in a single expression,
\begin{equation}
	[J_{\mu\nu},J_{\rho\sigma}] =-i (g_{\mu\rho}J_{\nu\sigma} + g_{\nu\sigma}J_{\mu\rho} - g_{\nu\rho}J_{\mu\sigma} - g_{\mu\sigma}J_{\nu\rho})\,. 
\end{equation}
These six generators form a sub-algebra. The subgroup generated by them is called the special Lorentz group, noted as $SO(3,1)\subset\mathcal{G}$. 
Besides, the 4 translation generators form a 4-vector, thus they commute to each other $[P_\mu,P_\nu]=0$, and the commutation with the other generators is
\begin{equation}
	[J_{\mu\nu},P_\rho] = -i(g_{\mu\rho}P_\nu - g_{\nu\rho}P_\mu) \,.
\end{equation}
The translation generators also form a sub-algebra, which is special since it is abelian and is closed under the commutation of a translation generator and any other \pg generator. Thus, this sub-algebra is an abelian ideal of the \pg group. 
The subgroup generated by the translation generators $P_\mu$ is the translation group $R^4$. Accordingly, the \pg group can be expressed as a semi-direct product of the translation group and the special Lorentz group,
\begin{equation}
	\mathcal{G} = R^4 \ltimes SO(3,1)\,.
\end{equation}
Thus, any \pg transformation can be noted as $(\Lambda,b)$, where $\Lambda\in SO(3,1)$ is called a Lorentz transformation, and $b\in R^4$ is the translation parameter. Its application on the coordinate $x^\mu$ is 
\begin{equation}
    (\Lambda,b) x^\mu = x'^\mu = \Lambda^\mu{}_\nu x^\nu + b^\nu\,.
\end{equation}
The application of two successive transformations is
\begin{equation}
    (\Lambda',b')(\Lambda,b) x^\mu = (\Lambda',b')x'^\mu = (\Lambda'\Lambda)^\mu{}_\nu x^\nu  + \Lambda'^\mu{}_\nu b^\nu + b'^\mu\,,
\end{equation}
thus we obtain the multiplication rule of the \pg group that
\begin{equation}
    (\Lambda',b')(\Lambda,b) = (\Lambda'\Lambda,\Lambda' b + b')\,.
\end{equation}




\paragraph{Induced Representation}

The dimension-4 Minkowski space furnishes a representation of the \pg group, which is called the fundamental representation, or the defining representation. It is not unitary since the boost generators are not Hermitian. As mentioned before, all the finite-dimensional representations of the non-compact groups are not unitary. Nevertheless, the unitary but infinite-dimensional representations exist. These representations are the so-called induced representation~\cite{Wigner:1939cj,Mackey1970,Weinberg:1995mt}, which is obtained from the irreducible representations of some subgroups.  

To discuss the unitary representations of the \pg group, we introduce the Casimir elements of the \pg group.
Since it is of rank 2, there are 2 Casimir elements. One is the inner product of the 4 translation generators, 
\begin{equation}
M^2 = P^2 = P_0^2-P_1^2-P_2^2-P_3^2\,,
\end{equation}
whose eigenvalues $m^2$ take different values for different irreducible representations of \pg group.
To find another Casimir element, we define a 4-vector
\begin{equation}
	W^\mu = \frac{1}{2}\epsilon^{\mu\nu\rho\sigma}J_{\nu\rho}P_\sigma\,,
\end{equation}
called the Pauli-Lubanski vector. It can be verified that $W^2 = W^\mu W_\mu$ commutes with all the \pg generators. 
Thus, we obtain the two Casimir elements of the \pg group $M^2,W^2$, and the irreducible representations are characterized by their eigenvalues $(m, w)$, of which a state vector is noted conventionally as
\begin{equation}
	|p,w\rangle \in (m,w)\,,\text{satisfying}\quad p^2 = m^2\,.
\end{equation}
The representation $(m,w)$ is of infinite dimension, since $p$ is continuous.
All these state vectors compose the Hilbert space of the \pg group.
These vectors can be interpreted as a single particle of momentum $p$, then the eigenvalue $m^2$ can be interpreted as the particle masses, thus $m$ is positive semi-definite, $m\geq 0$. For the induced representation, Different masses determine different subgroups to be considered, which are called little groups. 
$w$ is the eigenvalue of the Casimir element $W^2$ and is another quantum number of the particles, whose physical interpretation is also dependent on mass, and will be discussed next.


\paragraph{Little Group}

To illustrate the second Casimir element $W^2$, we consider some specific values of momentum $P$, which means we choose some special frames of reference. For a massive case, $P^2=m^2>0$, we can always choose the rest frame in which the spatial momentum components are zero,
\begin{equation}
P^* = (m,0,0,0)\,,
\end{equation}
where we have used $*$ to indicate it is not arbitrary. Given the fixed momentum $P^*$, the Pauli-Lubanski vector is reduced to 
\begin{equation}
	W^\mu \rightarrow W^i = \frac{m}{2}\epsilon^{ijk}J_{ij} = mJ^i\,,
\end{equation}
which are just the 3 spatial rotation generators. These generators are special since they do not change the fixed momentum $P^*$, $J_i P^* = P^*$. Thus, they generate a $SO(3)$ subgroup of \pg group under which the momentum $P^*$ is invariant. This group is called the little group of $P^*$. Consequently, the Casimir element $W^2$ of the massive particles is just the spin angular momentum,
\begin{equation}
	W^2 = -m^2J^2 = -m^2\sum_{i=1}^3J_i^2\,,
\end{equation}
its eigenvalues of an irreducible representation is $-m^2J(J+1)$ with $J$ half-integer. The one-particle state of momentum $p^*$ is noted as $|p^*,s,\sigma\rangle$, where $s$ is the spin characterizing the irreducible representations of the little group $SO(3)$, and $\sigma = -s,-s+1,\dots,s-1,s$ is the spin component along a specific axis. For a little group transformation $R\in SO(3)$, the momentum $p^*$ is unchanged, and the indices $\sigma$ transform irreducibly,
\begin{equation}
	|p^*,s,\sigma\rangle \rightarrow |p^*,s,\sigma'\rangle = \sum_{\sigma=-s}^{s}|p^*,s,\sigma\rangle D^{(s)}(R)_\sigma{}^{\sigma'}\,,
\end{equation}
where $D^{(s)}$ is the representative matrix.

Massless particles are different since a rest frame does not exist. We can choose another frame in which the spatial momentum is along the 3rd coordinate,
\begin{equation}
P^* = (1,0,0,1)\,.
\end{equation}
The resultant $W^\mu$ is
\begin{equation}
W^\mu = (J_3,J_1-K_2,J_2+K_1,J_3)\,,
\end{equation}
where there are 3 distinct generators 
\begin{equation}
W_1 = J_1-K_2\,,\quad W_2 = J_2+K_1\,,\quad W_3 = J_3\,.
\end{equation}
It is direct to verify these generators keep the fixed momentum $P^*$ unchanged, $W^iP^*=P^*$, thus they generate the little group of the massless momenta. To find out this group, we consider the algebra of the generators,
\begin{equation}
	[W_1,W_2]=0\,,\quad [W_3,W_1]=-W_2\,,\quad [W_3,W_2]=W_1\,.
\end{equation}
These furnish the Lie algebra of the 2-dimensional Euclidean group $E(2)$, the rotation and translation group of 2-dimensional space, where $W_3$ is responsible for the rotation of the 2-dimensional plane, while $W_1,W_2$ are responsible for the translations along two different directions. However, the two translations generate gauge transformations and have no observing effects, so we can consider only the generator $J_3$ and reduce the little group furthermore to the abelian $U(1)$~\cite{Zee:2016fuk}. The eigenvalues of $J_3$ referred to as helicity, are used to characterize the irreducible representations. 
Different from the massive case, all the irreducible representations of the $U(1)$ group are of dimension 1. Thus we can note the massless one-particle state as $|p^*,h\rangle$, and under the little group transformation $\exp(i\varphi)\in U(1)$, it transforms as
\begin{equation}
	|p^*,h\rangle \rightarrow e^{ih\varphi}|p^*,h\rangle\,.
\end{equation}

\paragraph{One-Particle State}

Now we move on from the fixed frame of reference to general frames. 
Suppose in a frame of reference, the particle momentum is $p$, which is related to the fixed one by a \pg transformation,
\begin{equation}
	p = L_p p^*\,,\quad L_p\in \mathcal{G}\,.
\end{equation}
Then we define the corresponding one-particle state as
\begin{align}
	\text{massive}:\quad & |p,s,\sigma\rangle\equiv U(L_p)|p^*,s,\sigma\rangle\,, \\
	\text{massless}:\quad & |p,h\rangle \equiv U(L_p)|p^*,h\rangle\,,
\end{align}
where $U(L_p)$ is the representative of $L_p$ in the Hilbert space. By this definition, we can obtain the transformations of a general one-particle state under the \pg group from the transformations under the little groups, where the latter have been obtained.
For convenience, we unify the spin and helicity indices by $I$, $\{s,\sigma\}/h \rightarrow I$, then a general one-particle state is noted as $|p,I\rangle$. For any $\Lambda\in\mathcal{G}$, it acting on the one-particle state as
\begin{equation}
|p,I\rangle \rightarrow U(\Lambda)|p,I\rangle\,,
\end{equation}
which corresponds to a state with momentum $p' = \Lambda p$. Such states are defined as $|p',I\rangle = U(L_{p'})|p^*,I\rangle$, thus we have
\begin{equation}
	U(\Lambda)|p,I\rangle = U(\Lambda)U(L_p)|p^*,I\rangle = U(L_{p'})U^{-1}(L_{p'})U(\Lambda)U(L_p)|p^*,I\rangle = U(L_{p'})U(L^{-1}_{p'}\Lambda L_p) |p^*,I\rangle\,, 
\end{equation}
Noting $\Lambda' = L^{-1}_{p'}\Lambda L_p$, it leaves the fixed momentum $p^*$ unchanged, $\Lambda' p^* = p^*$, so it is an element of little group.
The little group actions on the massless and massive states are known,
\begin{equation}
\label{eq:littlegroup}
	U(\Lambda')|p^*,I\rangle = \left\{\begin{array}{ll}
			|p^*,s,\sigma\rangle D^{(s)}(\Lambda')_\sigma{}^{\sigma'} & \text{massive} \\
			|p^*,h\rangle e^{ih\varphi} & \text{massless}
	\end{array}\right.
\end{equation}
Thus, the general transformation is 
\begin{equation}
	U(\Lambda)|p,I\rangle = \left\{ \begin{array}{ll}
			|p',s,\sigma\rangle D^{(s)}(\Lambda')_\sigma{}^{\sigma'} & \text{massive} \\
			|p',h\rangle e^{ih\varphi} & \text{massless}
\end{array}
\right.\,,\quad \Lambda \in \mathcal{G}\,,
\end{equation}
where the linearity of the representation has been used. 
Such representations obtained from the subgroup representations are called induced representations. In addition to the Lorentz transformations, the translations $e^{iP^\mu b_\mu}$ applying on the one-particle state gives 
\begin{equation}
    e^{iP^\mu b_\mu} |p,I\rangle = e^{ip^\mu b_\mu}|p,I\rangle\,.
\end{equation}
The one-particle states of specific mass and spin (helicity) compose a unitary representation, which is guaranteed by the unitarity of the little group representation in Eq.~\eqref{eq:littlegroup}.

\subsection{Discrete Symmetry}

In addition to the Lie groups, the discrete groups such as $Z_2$ also apply to particle physics, for example, the charge conjugation (C) is a discrete symmetry of elementary particles related to the gauge symmetry.
Letting $G$ be the gauge group, particle $|r\rangle$ transforms in the irreducible representation $r$ of $G$,
\begin{equation}
	|r\rangle \rightarrow D^{(r)}(g) |r\rangle\,,\quad g\in G\,.
\end{equation}
Charge conjugation transfers the particle to the anti-particle, $\overline{|r\rangle}$, which transforms in the complex conjugation of the representation $r$, noted as $r^*$,
\begin{equation}
	\label{eq:c1}
	|r\rangle \xrightarrow{C} \overline{|r\rangle} \rightarrow D^{(r)*}(g) \overline{|r\rangle}\,,\quad g\in G\,.
\end{equation}
In QCD and QED, the charge conjugation is conserved, but it is violated in the electroweak theory. Anyway, for the theories conserving C, the symmetry can be taken into account by a group-theoretical technique, automorphism~\cite{Henning:2021ctv,Graf:2020yxt,Bischer:2022rvf,Ohki:2023zsn}.

As mentioned before, an automorphism of a group $G$ is a bijection mapping $G$ to itself. All the automorphisms form a group noted as $\text{Aut}(G)$. A special case is the adjoint map,
\begin{equation}
	\text{Ad}_h : g\rightarrow h g h^{-1}\,,\quad h,g\in G\,,
\end{equation}
which are alled inner automorphism, and form a subgroup $\text{Inn}(G) \subset \text{Aut}(G)$. All the automorphisms other than the inner ones are called the outer automorphisms, which form the quotient group that
\begin{equation}
	\text{Out}(G) = \text{Aut}(G)/\text{Inn}(G)\,.
\end{equation}

The charge conjugation induces an automorphism of the gauge group $G$ according to Eq.~\eqref{eq:c1}, which can be expressed by the representation-independent form,
\begin{equation}
	\text{C}:\quad g \rightarrow g^*\,.
\end{equation}
It follows immediately that the charge conjugation furnishes a $Z_2$ group, and a natural question is whether the C is an inner or outer automorphism.

The answer about the special unitary groups $SU(N)\,,N\geq 2$ is simple. If it is inner, we could find a group element $h\in SU(N)$ satisfying $hgh^\dagger = g^*$, or 
\begin{equation}
	\label{eq:schur}
	(h^*h) g = g (h^*h)\,,
\end{equation}
which means $(h^*h)$ commutes with all $g\in G$. Since the result is representation-independent, we focus on a specific irreducible representation $r$. According to Schur lemma, Eq.~\eqref{eq:schur} imples $D^{(r)*}(h)D^{(r)}(h)\propto I_{d_r}$. Considering $SU(2)$ and its fundamental representation, we can find such an element,
\begin{equation}
	h=i\sigma^2 = \left(\begin{array}{cc}
			0 & 1 \\ -1 & 0
	\end{array}\right) \in SU(2)\,,\text{satisfying }h^*h=-I_2\,. 
\end{equation}
Nevertheless, such elements do not exist in the group $SU(N)$ of $N\geq 3$.
Thus, we conclude that the charge conjugation is an inner automorphism of the $SU(2)$ group, and an outer automorphism of $SU(N)\,,N\geq 3$.

If the charge conjugation is an inner automorphism, the particles and anti-particles are equivalent, which means they can be transferred to each other by a similar matrix, for example, $i\sigma^2$ of the $SU(2)$ fundamental representation. While if the charge conjugation is outer, the particles and anti-particles are distinguished.

As an example, we consider the standard model of gauge group $SU(3)_c\times SU(2)_L\times U(1)_Y$. The quarks are $SU(3)_c$-triplets, thus the quarks and anti-quarks are different. As for the leptons, they are $SU(3)_c$ singlets and $SU(2)$-doublets. Although the charge conjugation of the $SU(2)$ group is inner, the leptons and anti-leptons are also different, since they are also charged under the $U(1)_Y$ group. Since $U(1)_Y$ is abelian, all the automorphisms are outer.

Besides, the right-handed neutrinos $\nu_R$ are another example of the special $SU(2)$. Although $\nu_R$ is absent in the standard model because it is trivial under the gauge group, it is an example that the particles and anti-particles are equivalent. As a spinor, $\nu_R$ is of the $SU(2)$ fundamental representation. Since the charge conjugation is an inner automorphism, we can construct an invariant $\nu_R^T i\sigma^2 \nu_R$. Such a bilinear term contributes to the mass of the neutrino, which, distinguished from the Dirac mass, is called the Majorana mass~\cite{Majorana:1937vz}. Although the masses of the neutrinos have been verified by oscillation experiments~\cite{ParticleDataGroup:2024cfk}, whether it is a Majorana or a Dirac fermion is still unknown. This is expected to be determined by the non-oscillation experiments, such as the neutrinoless double beta decay in the future~\cite{Avignone:2007fu,Cirigliano:2022oqy}.

\section{Gauge Theory}\label{gauge}

\newcommand{\ii}{\mathrm{i}}
\newcommand{\dd}{\mathrm{d}}

With the language of group theory in hand, we can now articulate the gauge principle, the most fruitful concept in modern field theory. 

\subsection{Local Symmetry and Gauge Invariance}
The global symmetry introduced above asserts that only the phase differences are observables. It implies that an experimenter in a lab on Earth and another in a distant galaxy must somehow agree on their choice of phase convention for the same field in order to describe the phenomenon of interference. This is analogous to the case of Newtonian mechanics, where there is a sense of absolute time synchronized at different space points. However, the principle of relativity suggests that the notion of time should depend on observers and, in the same spirit, so does the convention of phases. Therefore, the laws of physics must be formulated so that they are independent of these arbitrary local choices. In particular, the transformation parameter $\alpha$ of the symmetry group should be promoted to a spacetime-dependent function $\alpha(x)$, while the theory should remain the same under this local ``gauge'' transformation \cite{Weyl1929Elektron,Fock1926ZPhy}
\begin{equation}
	\psi(x) \to \e^{\ii\alpha(x)}\psi(x)
\end{equation}

The theory is easily invariant under such a local transformation if there are no derivative terms in the Lagrangian. But for a field to have dynamics, it must have at least a kinetic term that involves a spacetime derivative, which would recognize the phase difference between the neighboring points. 
To see how this works, consider a fermion field $\psi(x)$ transforms under $U(1)$ as $\psi(x) \to \e^{\ii\alpha}\psi(x)$. Promoting this to a local symmetry $\psi(x) \to \e^{\ii\alpha(x)}\psi(x)$ breaks the invariance of its kinetic term $\ii\bar{\psi}\gamma^\mu \partial_\mu \psi$. To cure this, one introduces a gauge field $A_\mu(x)$ which transforms inhomogeneously as $A_\mu \to A_\mu - \partial_\mu \alpha$ and replaces the ordinary derivative with the ``covariant derivative'' 
\begin{equation}
D_\mu = \partial_\mu + \ii A_\mu \ . 
\end{equation}
The covariance of this operator is indicated by the homogeneous transformation
\begin{equation}
D_\mu\psi(x) \to \e^{\ii\alpha(x)} D_\mu\psi(x) \ , 
\end{equation}
which guarantees the gauge invariance of the kinetic term and many other derivative couplings. Therefore, the kinetic term of the fermion charged under a gauge $U(1)$ is written as
\begin{equation} 
\ii\bar\psi\gamma^\mu D_\mu\psi = \ii\bar\psi\gamma^\mu\partial_\mu\psi - \bar\psi\gamma^\mu A_\mu\psi \ , 
\end{equation}
where the second term is called the minimal gauge coupling of the fermion $\psi$, directly resulting from replacing the partial derivative in the kinetic term with the covariant derivative. In general, the source of the gauge field in the minimal gauge coupling is exactly the Noether current of the corresponding global symmetry, such as $J^\mu=\bar\psi\gamma^\mu\psi$.
For non-Abelian groups, where the parameter $\alpha(x) = \alpha^a(x) T^a$ is a matrix and $T^a$ are the generators of the Lie algebra, the gauge field $A_\mu(x) = A_\mu^a(x)T^a$ is naturally a matrix field as well, with gauge transformation \cite{Yang:1954ek,PhysRev.101.1597}
\begin{equation}\label{eq:gauge_A}
    A_\mu \to A_\mu^{(\alpha)} = \e^{\ii\alpha}(A_\mu-\ii\partial_\mu)\e^{-\ii\alpha} = A_\mu - D_\mu \alpha + O(\alpha^2) \ .
\end{equation}
Since $\alpha(x)$ is in the adjoint representation, its covariant derivative takes the form $D_\mu\alpha = \partial_\mu\alpha + \ii [A_\mu,\alpha]$.

The introduction of covariant derivative and gauge field is similar as in the case of general relativity, where the local spacetime symmetry (diffeomorphism) suggests a covariant derivative that involves a Christoffel symbol $\nabla_\mu=\partial_\mu+\Gamma_\mu$. In mathematics, both of them are called a connection on the fiber bundle \cite{Wu:1975es,Nakahara:2003nw}, where an algebraic structure called a typical fiber, which is the coordinate frame in the case of general relativity and the representation space in the case of gauge theory, grows on the spacetime base manifold. The gauge symmetry corresponds to the independent choice of coordinates on the different fibers, which demands a connection 1-form $\omega_\mu$ on the bundle structure that defines the parallel transport across the fiber bundle via the associated covariant derivative $D_\mu$. 
In both cases, there is a curvature 2-form on the fiber bundle, which measures the obstruction to commuting parallel transports in different directions. It manifests as the Riemann curvature $R$ in general relativity and as the field strength tensor $F_{\mu\nu}$ in the gauge theory. These are non-zero effects of the connection that cannot be eliminated by a local transformation, and are physically observable. This is exactly why we can observe the electric and magnetic fields, which are components of the field strength tensor. 
In particular, it can be defined as the commutator of the covariant derivatives $[D_\mu,D_\nu] = \ii F_{\mu\nu}$, or 
\begin{equation}
    F_{\mu\nu} = \partial_\mu A_\nu - \partial_\nu A_\mu + \ii[A_\mu,A_\nu] \ , 
\end{equation}
where the last term exists for non-Abelian groups where $A_\mu = A_\mu^a T^a$ are matrices and is crucial for the self-interacting feature of the non-Abelian gauge field. In terms of the component fields $A^a_\mu$, it can be written as
\begin{equation}
    F^a_{\mu\nu} = \partial_\mu A^a_\nu - \partial_\nu A^a_\mu - f^{abc}A^b_\mu A^c_\nu \ .
\end{equation}

The dynamics of the gauge field should be given by a quadratic term with derivatives in the Lagrangian. In four spacetime dimensions, a natural choice is the operator $F_{\mu\nu}F^{\mu\nu}$, which reproduces the renowned Maxwell’s equations as its equation of motion
\begin{equation}
    D^\nu F_{\mu\nu} = J_\mu\ .
\end{equation}
Another quadratic operator is the CP odd $F_{\mu\nu}\tilde{F}^{\mu\nu}$, which turns out to be a total derivative and serves as a topological term
\begin{equation}
    \tr(F_{\mu\nu}\tilde{F}^{\mu\nu}) \equiv \epsilon^{\mu\nu\rho\lambda}\tr(F_{\mu\nu}F_{\rho\lambda}) = 2\epsilon^{\mu\nu\rho\lambda}\partial_\mu \tr\left(A_\nu \partial_\rho A_\lambda - \frac{2\ii g}{3}A_\nu A_\rho A_\lambda \right) \ .
\end{equation}
It does not affect the dynamics of the gauge field alone, but would manifest in loop-level processes and contribute to the neutron electric-dipole-moment (EDM). Its negligible value as measured in the experiment is a well-known puzzle of the Standard Model, named the strong CP problem. 
In sum, the gauge-invariant classical Lagrangian of a gauge theory has the following form
\begin{equation}\label{eq:Lag_0}
    \mathcal{L} = -\frac{1}{4}\sum_aF^a_{\mu\nu}F^{a\mu\nu} + \sum_f \ii\bar\psi_f \gamma^\mu D_\mu \psi_f + \sum_s |D_\mu\phi_s|^2
\end{equation} 
where the kinetic term is canonically normalized by a rescaling $A_\mu \to gA_\mu$ and the covariant derivatives are generally defined as $D_\mu=\partial_\mu + ig\sum_a A^a_\mu(x) T^a_{\mathbf{R}}$ with the generators $T^a_{\mathbf{R}}$ determined by the representation $\mathbf{R}$ of the field that it acts on. In particular, $T_{\mathbf{R}}=q$ is the charge of the field for a $U(1)$ gauge group.

\subsection{Is gauge symmetry a symmetry?}

While the classical Lagrangian is perfect for deriving the equations of motion for both the gauge fields and the matter fields, it has a big problem with quantization, which precisely stems from the gauge symmetry. Although we call it a 'symmetry', gauge symmetry is not a physical one. A physical symmetry relates distinct quantum states that form a representation space of the symmetry group. In contrast, gauge symmetry acts on field configurations without changing the physical quantum state. It merely represents a redundancy in our field theory description.
To demonstrate the point, we construct the gauge transformation on the Hilbert space.
Being a local symmetry with a spacetime dependent parameter $\alpha(x)$, the associated Noether current is \cite{Noether_1971}
\begin{equation}
    J^\mu(x) = \partial_\nu \left(F^{\nu\mu}(x)\alpha(x)\right)
\end{equation}
with a conserved Noether charge
\begin{equation} 
Q[\alpha(x)] = \int\Diff{x}{3} \nabla\cdot(\mathbf{E}(x)\alpha(x)) = \oint_{S_2} \dd\mathbf{\Sigma} \cdot \mathbf{E}(x) \alpha(x) 
\end{equation}
defined as an integration at spatial infinity $S^2$.
The global symmetry amounts to a constant $\alpha(x)=\alpha$, so that the Noether charge is the electric charge operator $Q_{\rm el} = \oint_{S_2} \dd\mathbf{\Sigma} \cdot \mathbf{E}(x)$ whose conservation imposes, according to the Noether's second theorem, a non-local constraint on the equations of motion for the gauge fields. 
It defines a transformation on the Hilbert space 
\begin{equation}\label{eq:global}
    \mc{U}_{\rm global}(\alpha) = \exp(\ii\alpha Q_{\rm el})
\end{equation}
In contrast, for a gauge transformation with $\lim_{x\to\infty}\alpha(x)=0$, we have trivially $Q[\alpha]\equiv 0$ which does not qualify as a Noether charge, and the induced transformation on the Hilbert space is the identity 
\begin{equation}
    \mc{U}_{\rm gauge}[\alpha(x)]= \exp(\ii Q[\alpha]) = \mathbbm{1} \ ,
\end{equation} 
indicating that the transformation does not change the physical quantum state at all $\mc{U}_{\rm gauge}|\psi\rangle_{\rm phy}=|\psi\rangle_{\rm phy}$.
In other words, the existence of gauge symmetry is reflecting a redundancy in our mathematical description of the field theory, instead of a fundamental property of nature.

We have seen from the above that not all gauge transformations are redundant. 
Besides the global transformation \eqref{eq:global}, there are actually an infinite number of physical symmetries with non-vanishing $\alpha(\mathbf{n})$ at the asymptotic spatial infinity, where $\mathbf{n}$ is the spatial direction on $S^2$.
The conservation of the corresponding charges $Q[\alpha(\mathbf{n})]$ has a direct and spectacular consequence: it implies the existence of an infinite number of Ward identities that the S-matrix must obey. These Ward identities are not empty; they are physically equivalent to the soft photon theorem. The theorem states that the amplitude for emitting a low-energy (soft) photon with polarization $\epsilon_\mu$ and momentum $k$ in a scattering process is universally given by \cite{Weinberg:1965nx}:
\begin{equation}
    \lim_{k\to 0}\mc{M}_{n+1}(\{\Phi\}_n;\gamma(k,\varepsilon)) = \sum_i q_i \frac{p_i\cdot\varepsilon}{p_i\cdot k}\,\mc{M}_n(\{\Phi\}_n) \ .
\end{equation}
where $q_i$ are the charges of the external particles. This factorization formula was long known from explicit Feynman diagram calculations. The revolutionary understanding is that this soft theorem is the Ward identity of the asymptotic symmetry $Q[\alpha(\mathbf{n})]$. The soft photon itself is the Goldstone boson associated with the spontaneous breaking of these asymptotic symmetries by the vacuum.

This triad of concepts — Asymptotic Symmetries, Soft Theorems, and Memory Effects (a classical, measurable imprint left by passing radiation)—forms a cornerstone of modern theoretical physics, often called the Infrared Triangle \cite{Strominger:2013jfa}. It reveals that what was once considered a mere technical redundancy in the interior of spacetime encodes a rich, physical symmetry structure at its boundary, governing the long-range dynamics and imposing powerful constraints on the scattering amplitudes.

\subsection{The quantization of gauge theory}

Dirac's canonical quantization of constrained systems \cite{Dirac_1950} reveals that gauge redundancies manifest as first-class constraints within the Hamiltonian formalism. For the gauge field theory, there is a primary constraint $\partial\mc{L}/\partial \dot{A}_0 \equiv \mathbf{\pi}^0\approx0$ and a secondary constraint $[\mathbf{\pi}^0,H]=\nabla\cdot E\approx0$, where $\approx$ denotes the weak equality, which holds on the subspace of phase space defined by the constraints. These constraints, which generate the gauge transformations in eq.~\eqref{eq:gauge_A}, render the naive Poisson bracket algebra inconsistent and the phase space propagator singular, as it attempts to quantize physically equivalent field configurations as distinct degrees of freedom. 
Consequently, one must either employ Dirac's method of imposing constraints on physical states or explicitly eliminate the redundancy by reducing the phase space through a gauge fixing procedure. However, this reduction often sacrifices manifest Lorentz invariance, which is crucial for maintaining the elegance and calculational power of relativistic field theory.

This tension between a covariant formalism and a well-defined quantization procedure motivates the Faddeev-Popov (FP) procedure \cite{Faddeev:1967fc}. Instead of directly solving the constraints, the FP method cleverly factors the infinite volume of gauge orbits out of the path integral. The naive path integral of gauge fields wildly overcounts the physically identical configurations related by gauge transformation 
\begin{equation}
    \mathcal{Z}[A] \sim \int\mathcal{D}A_\mu \exp\left[\ii\int\dd^4x \,\mc{L}(x) \right]
\end{equation}
where $\mc{L}$ takes the form in eq.~\eqref{eq:Lag_0}. To reduce the gauge redundancy in the integration measure, we introduce the gauge fixing condition via a delta function $\delta(G(A))$. It integrates to $1$ on the group manifold as
\begin{equation}
    1=\int \mathcal{D}\alpha\,\delta(G(A^{(\alpha)})) \Delta_{\rm FP}[A^{(\alpha)}]
\end{equation}
where the $\mathcal{D}\alpha$ is the Haar measure of the gauge group and $A^{(\alpha)}$ is the gauge transformed field given in eq.~\eqref{eq:gauge_A}, which has the infinitesimal version $\delta A^{a}_\mu = -(\delta^{ab}\partial_\mu +f^{abc}A^c_{\mu})\delta\alpha^b \equiv -D_\mu^{ab}\delta\alpha^b$. The Jacobian is called the Faddeev-Popov determinant
\begin{equation}
    \Delta_{\rm FP}[A^{(\alpha)}]=\det(M) \ ,\qquad  M^{ab}=\frac{\delta G^a(A^{(\alpha)}(x))}{\delta \alpha^b(y) }
    =-\frac{\partial G^a}{\partial A^b_\mu} D^{bc}_\mu\delta^{(4)}(x-y)\ ,
\end{equation}
It is convenient to take $G^a=\partial^\mu A_\mu^a(x)-\omega^a(x)$, so that for $\omega=0$ the delta function exactly gives the Lorentz condition $\partial^\mu A^a_\mu=0$. Now we have $M^{ab}(x,y)=-\partial^\mu D_\mu^{ab}\delta^{(4)}(x-y)$.
The determinant is then written as a Gaussian integral of a fermionic auxiliary field, the ghost $(c,\bar{c})$, as
\begin{equation}
    \Delta_{\rm FP}[A^{(\alpha)}]= \exp\left[\ii\int\dd^4 x\, \mathcal{L}_{\rm gh} \right] \ ,\quad \mathcal{L}_{\rm gh}=\int\dd^4 y\,\bar{c}^a(x) M^{ab}(x,y) c^b(y)
    =-\left(\partial^\mu\bar{c}^a \partial_\mu c^a - gf^{abc}A^a_\mu \partial^\mu\bar{c}^b c^c\right)
\end{equation}
Finally, since the path integral is independent of $\omega^a$ in the function $G^a(x)$, we may multiply it by a numerical constant from an auxiliary $\omega$ integral and get rid of the delta function
\begin{equation}
    \int\mathcal{D}\omega\, \exp\left[-\frac{\ii}{2\xi}\int\dd^4 x\, \omega^a\omega^a \right] \delta(\partial^\mu A_\mu^a - \omega^a)
    = \exp\left[\ii\int\dd^4 x\, \mathcal{L}_{\rm gf} \right] \ ,\quad \mathcal{L}_{\rm gf}=-\frac{1}{2\xi}(\partial^\mu A_\mu)^2
\end{equation}
Together, we get a workable gauge-fixed action, with a redundant integration of the parameter $\alpha$ factored out
\begin{equation}
    \int\mathcal{D}\omega\, \exp\left[-\frac{\ii}{2\xi}\int\dd^4 x\, \omega^a\omega^a \right] \times \mc{Z}[A] = \int \mathcal{D}\alpha \times \int\mathcal{D}A_\mu \exp\left[\ii\int\dd^4x \,\left(\mc{L} + \mc{L}_{\rm gf} + \mc{L}_{\rm gh}\right) \right]
\end{equation}


In fact, the Faddeev-Popov procedure does not really fix the gauge when it introduces the weighted average in $\omega$. Instead of reducing the redundant degrees of freedom, it leaves all the components of $A_\mu$ in the path integral, and even extends the Hilbert space by introducing the ghost/anti-ghost fields. What it actually does is to effectively eliminate the singularity in the propagator
\begin{equation}
    \mc{L} + \mc{L}_{\rm gf} = \frac{1}{2}A_{\mu} \left(g^{\mu\nu}\Box - (1-\xi^{-1})\partial^\mu\partial^\nu\right)A_\nu\qquad\Rightarrow\qquad D_{\mu\nu}(k) = \frac{-\ii}{k^2+i\epsilon}\left[g^{\mu\nu}-(1-\xi)\frac{k^\mu k^\nu}{k^2}\right] \ .
\end{equation}
The gauge invariance of the new action is guaranteed by a global BRST symmetry developed by Becchi, Rouet, Stora \cite{Becchi:1974md,Becchi:1975nq}, and Tyutin \cite{Tyutin:1975qk}. 
It is a fermionic symmetry acting on the gauge fields and the ghosts as
\begin{equation}
    \delta_{\rm B}A_\mu = D_\mu c\ ,\qquad \delta_{\rm B}c = \frac{\ii g}{2}[c,c]\ ,\qquad \delta_{\rm B}\bar{c}=\ii B \ ,\qquad \delta_{\rm B}B=0\ ,
\end{equation}
where $B$ is the Lautrup-Nakanishi auxiliary field. It is immediately observed that the transformation of the gauge field is a special gauge transformation with parameter $\alpha = -c$, hence the gauge invariant Lagrangian $\mc{L}$ is automatically BRST invariant. The above transformation rules are demanded by a requirement of its nilpotency $\delta_{\rm B}^2 = 0$.
In terms of the BRST transformation, the whole Faddeev-Popov procedure can be written as
\begin{equation}
    \mc{L} \to \mc{L} + \delta_{\rm B}\left[\bar{c}^a \big(\frac{\xi}{2}B^a - \partial^\mu A_\mu^a\big)\right]  = \mc{L}+\mc{L}_{\rm gf} + \mc{L}_{\rm gh}\ ,
\end{equation}
where the nilpotency guarantees the BRST-invariance of the second term, and thus proves the BRST symmetry of the gauge-fixed Lagrangian.
The BRST symmetry induces a conserved BRST charge $Q_{\rm B}$, which inherits the nilpotency of the transformation $Q_{\rm B}^2=0$. 
This property allows for a cohomological definition of the physical Hilbert space as
\begin{equation} 
	\mathcal{H}_{\rm phy} = \frac{\operatorname{Ker} Q_{\rm B} }{ \operatorname{Im} Q_{\rm B} } \ ,
\end{equation}
which means that physical states $\ket{\psi}$ should be BRST-closed $Q_{\rm B}\ket{\psi}=0$, while the BRST-exact states $\ket{\psi} = Q_{\rm B}\ket{\chi}$ have null norms and should be quotiented out. 
This cohomology space is invariant under a unitary and BRST-symmetric time evolution; under a careful analysis, one could confirm that among all the polarizations and ghosts, only the two transverse polarizations are in the cohomology. Therefore, it is guaranteed that no unphysical states can be produced in the scattering process, which is the manifestation of the gauge invariance.
The mathematical description of this property is the Slavnov-Taylor (ST) identity \cite{Slavnov:1972fg,Taylor:1971ff} of the quantum effective action $\Gamma[A_\mu,\dots]$, which is basically the Ward-Takahashi identity of the BRST symmetry
\begin{equation}
    \mc{S}(\Gamma) \equiv \int\dd x \sum_\Phi\frac{\partial \Gamma}{\partial \Phi(x)} \vev{\delta_{\rm B}\Phi(x)\dots} = 0
\end{equation}
This identity severely restricts the form of possible ultraviolet counterterms and confines all infinities to a redefinition of a finite number of physical parameters and field normalizations, proving the renormalizability of gauge theories \cite{tHooft:1971qjg}. Therefore, the consistency of this entire quantized framework relies on the BRST symmetry being non-anomalous at the quantum level, which leads us to the crucial topic of gauge anomaly cancellation.

\subsection{Cancellation of gauge anomaly}

The symmetry of the classical action could be violated in a non-perturbative way by quantum anomaly \cite{PhysRev.177.2426,bell1969pcac,Fujikawa:1979ay}. It becomes especially important for gauge theory, as the breaking of gauge symmetry could be a disaster for the whole theory \cite{Bouchiat:1972iq}. Unphysical degrees of freedom in the gauge field (the longitudinal and temporal polarizations) would become non-decoupling due to the breakdown of the ST identity, which then violate the crucial principle of unitarity. Therefore, unlike the anomaly for global symmetry that is acceptable as an important feature, the gauge anomaly must be canceled in a physically valid quantum theory.

The gauge anomaly comes from the chiral fermion loops, proportional to the symmetric traces $D^{abc}\equiv \tr\left[T^a \{T^b,T^c\}\right]$ where $T^a$ are the generators of some gauge group in the representation of the chiral fermion. For example, the Standard Model gauge anomaly can be listed as follows (take the left-handed fermions; for right-handed fermions, take their charge conjugate)
\begin{align}
    SU(3)_C\times SU(3)_C\times SU(3)_C: && &\sum_{f\in \mathbf{3}} d^{abc} - \sum_{f\in \bar{\mathbf{3}}} d^{abc}=0 \ ,\\
    SU(3)_C\times SU(3)_C\times U(1)_Y: && &\sum_{f}Y_fT(\mathbf{R}_f)\delta^{ab} =0\ , \\
    SU(2)_L\times SU(2)_L\times U(1)_Y: && &\sum_{f}Y_f T(\mathbf{R}_f)\delta^{ij} =0\ , \\
    U(1)_Y\times U(1)_Y\times U(1)_Y: && &\sum_{f} Y_f^3 =0\ .
\end{align}
Another condition that is usually listed together is the gravity$^2\times U(1)_Y$ anomaly, which has two gravitons in the triangle that universally couple to all the fermions, thus the condition is 
\begin{equation}
    \sum_f Y_f =0 \ .
\end{equation}
All the fermions in the Standard Model are chiral: $f\in\{Q, u^c, d^c, L, e^c\}$, but somehow they satisfy all the gauge anomaly cancellation conditions. It is often considered a strong and highly non-trivial evidence for a Grand Unified Theory (GUT), where there is only a simple gauge group and hence much less gauge anomaly conditions. Due to the 't Hooft's anomaly matching condition \cite{Hooft1980}, the Standard Model would be anomaly-free if it comes from an anomaly-free GUT. For example, in the $SU(5)$ GUT \cite{PhysRevLett.32.438}, the above 5 conditions merge into a single one $SU(5)^3$, which is satisfied by the fermion representation $\bar{\mathbf{5}}\oplus \mathbf{10}$; for $SO(10)$ GUT \cite{FRITZSCH1975193}, it is even more natural because all its representations are anomaly free. 

The gauge anomaly can also be understood from the ST identity \cite{Wess:1971yu,Stora:1983ct,Zumino:1983rz}. The triangle loop of chiral fermions could in principle generate non-zero $\mc{S}(\Gamma)$: if it is still BRST-exact $\mc{S}(\Gamma) = \delta_{\rm B}\Gamma'$, it could be absorbed by a counterterm $-\Gamma'$ in the action; however, if it's in the cohomology, it is a real anomaly that cannot be canceled by counterterms, and would break the gauge symmetry. To see why it breaks the symmetry, we have
\begin{equation}
    \mc{S}(\Gamma) \supset \int\dd x \frac{\partial \Gamma}{\partial A_\mu(x)} \vev{\delta_{\rm B}A_\mu(x)\dots} \supset \int\dd x \left\langle J^\mu(x)\dots\right\rangle \vev{D_\mu c(x)\dots} \quad\Rightarrow\quad \frac{\delta}{\delta c}\mc{S}(\Gamma) \sim \vev{\partial_\mu J^\mu \dots} 
\end{equation}
The ONLY candidate for $\mc{S}(\Gamma)$, which should be a local operator with ghost number $+1$ and not BRST exact, is
\begin{equation}
    \mc{S}(\Gamma) \sim d^{abc} \int \dd^4 x\,c^a(x) F^b_{\mu\nu}(x)\tilde{F}^{c\mu\nu}(x) \ ,
\end{equation}
which precisely leads to the anomaly we expect $\vev{\partial_\mu J^{a\mu} \dots} \sim d^{abc}F^b_{\mu\nu}(x)\tilde{F}^{c\mu\nu}(x)$.

\subsection{The On-Shell Scattering Amplitude Approach}
\label{subsec:on-shell-approach}

Having detailed the challenges of gauge redundancy in the traditional Lagrangian formulation, we now turn to a modern alternative that circumvents this issue entirely: the on-shell scattering amplitude approach. 

It was then discovered by Parke and Taylor in \cite{Parke:1986gb,mangano1991multi} that some on-shell amplitudes in the gauge theory have an extremely simple expression in terms of the on-shell variables, despite the lengthy Feynman diagram computations due to the huge redundancy in the field theory and its Feynman rules. 
It reminds people of the Heisenberg's S-matrix program \cite{Heisenberg:1943}, which is deeply influenced by the operationalist philosophy of "using only observable quantities". He argued that since we can only measure initial and final states in scattering experiments, the theory should be formulated directly in terms of the S-Matrix (the operator that maps initial states to final states), bypassing the unobservable field operators and the problematic notion of local interactions. 
Although he failed due to the lack of theoretical preparations at the time, the idea reincarnate in the modern era, particularly for a non-redundant description of the gauge theories \cite{witten2004perturbative,Britto:2005fq,arkani2008tree}. By never introducing off-shell fields or a local Lagrangian, one avoids the gauge redundancy problem from the outset. 
The methodology is built upon three pillars:

\begin{enumerate}
    \item \textbf{Lorentz Invariance and Little Group Scaling:} The amplitude must be a Lorentz invariant function. For massless particles, this implies a specific scaling behavior under the ``little group''. In the spinor-helicity formalism, where a null momentum is written as \(p_{\alpha\dot{\alpha}} = \lambda_\alpha \tilde{\lambda}_{\dot{\alpha}}\), a helicity-\(h\) state under a little group transformation scales as \((\lambda, \tilde{\lambda}) \to (t\lambda, t^{-1}\tilde{\lambda})\) and the amplitude must satisfy
    \begin{equation}
        A(\{t_i \lambda_i, t_i^{-1} \tilde{\lambda}_i, h_i\}) = \left(\prod_i t_i^{-2h_i}\right) A(\{\lambda_i, \tilde{\lambda}_i, h_i\}).
    \end{equation}
    with all the spinor indices contracted to guarantee Lorentz invariance. The spinor contractions are usually denoted by
    \begin{equation}
        \epsilon^{\alpha\beta}\lambda_{i\alpha}\lambda_{j\beta} \equiv \vev{ij} \ ,\qquad \epsilon_{\dot\alpha\dot\beta}\tilde\lambda_i^{\dot\alpha}\tilde\lambda_j^{\dot\beta} \equiv [ij] \ .
    \end{equation}
    
    \item \textbf{Locality:} The amplitude is constructed to be a rational function of Lorentz invariants $\vev{ij},[ij]$, with its analytic structure dictated by physics. Simple poles occur only at Mandelstam invariants $s_I$ corresponding to the propagation of physical, on-shell intermediate states. 
    
    \item \textbf{Unitarity:} The residue at each pole is demanded by unitarity to factorize into a product of lower-point amplitudes:
    \begin{equation}
        \text{Res}_{s_I\to m^2} A_n = \sum_h A_L(\dots,h) \times A_R(-h,\dots).
    \end{equation}
    where $m$ and $h$ are the pole mass and the helicity of the intermediate state.
\end{enumerate}

As a canonical example, consider the three-gluon amplitude. Imposing locality, little group scaling, and the correct mass dimension uniquely fixes its form for, say, the mostly-minus helicity configuration:
\begin{equation}
    A(1^{a-},2^{b-},3^{c+}) = gf^{abc}\frac{\langle 12\rangle^3}{\langle 23\rangle\langle 31\rangle}.
\end{equation}
This expression is remarkably compact and contains no gauge redundancy; it is a function of the physical, on-shell spinors of the external particles. It is not derived from a Feynman rule but is \emph{constructed} as the simplest object consistent with the physical principles above. Yet, when expanded, it reproduces the complex structure of the Yang-Mills three-vertex. 
Furthermore, when we use the unitarity to construct the four-gluon amplitude, the consistent factorization in the various channels demands the Jacobi identity of the structure constant
\begin{equation}
    f^{abe}f^{cde} + f^{ace}f^{dbe} + f^{ade}f^{bce} = 0 \ ,
\end{equation}
requiring the interacting gluons to form a Lie algebra.
In this approach, \textbf{gauge invariance is not an input but an output}, manifesting itself through the decoupling of unphysical states, which is guaranteed by the on-shell construction. 

The elimination of the gauge redundancy can also be tracked by a hybrid method: using the spinor-helicity variables to express the Feynman rule calculation. The expression for the physical polarization vectors is 
\begin{equation}
	\varepsilon_+^\mu = \frac{\< r|\sigma^\mu|k]}{\sqrt{2}\<rk\>}\ ,\qquad
	\varepsilon_-^\mu = \frac{[r|\bar\sigma^\mu|k\>}{\sqrt{2}[rk]}\ .
\end{equation}
where $\ket{r},|r]$ are the reference spinors. The freedom of choosing the reference spinors reflects the gauge redundancy of the description in terms of the polarization vectors. However, the gauge-invariant on-shell amplitudes, in the end, would not depend on the reference spinors. One may confirm explicitly that they can be eliminated in the full amplitude calculation when all the Feynman diagrams are summed up. The on-shell construction simply skips the Feynman diagrams and tries to get the final result directly under the constraints of locality and unitarity, and the reference spinors do not show up throughout the process. Powerful tools like on-shell recursion relations (e.g., BCFW \cite{Britto:2005fq}) allow for the systematic construction of all higher-point tree-level amplitudes from the fundamental three-point amplitude, building the entire tree-level \(S\)-matrix from the ground up. A review on this whole topic can be found in \cite{Dixon:1996wi,henn2014scattering,Elvang:2015rqa,cheung2016tasi}

While conceptually elegant, the on-shell approach has its own subtleties. Establishing the existence and uniqueness of amplitudes for general multiplicity is non-trivial \cite{Elvang:2015rqa,Cheung:2017pzi,Cohen:2010mi}, and the formalism provides no direct access to off-shell information, such as correlation functions or background field dynamics. Despite these limitations, its focus on physical observables has led to profound insights into the structure of gauge theories themselves. 
A prime example is the Bern-Carrasco-Johansson (BCJ) color-kinematic duality, which reveals a deep-seated relationship between the kinematic and color structures of Yang-Mills amplitudes \cite{Bern:2008qj,Bern:2010ue}. This not only provides powerful computational tools but also leads to the double-copy construction, suggesting that gravity amplitudes can be constructed as a "square" of gauge theory ones \cite{Bern:2010ue}, hinting at a unified underlying kinematic algebra. 
Furthermore, the on-shell perspective was instrumental in the development of scattering equations and the Cachazo-He-Yuan (CHY) formalism \cite{Cachazo:2013hca,Cachazo:2013iea,Cachazo:2014nsa}, which provides a unified description of amplitudes in gauge theory, gravity, and scalars. These frameworks achieve a complete description of scattering processes without ever invoking the principle of gauge invariance, demonstrating that it is indeed an emergent consequence of more fundamental physical and mathematical constraints \cite{Bern:2019prr}.


\section{Symmetries of Standard Model}

The standard model (SM) of the elementary particles is an appropriate example of the application of gauge theory. Its gauge group is a direct product 
\begin{equation}
	\mathcal{G}_{\text{SM}} = SU(3)_C\otimes SU(2)_L \otimes U(1)_Y\,,
\end{equation}
in which the $SU(3)_C$ is the color group responsible for strong interaction, the $SU(2)_L\otimes U(1)_Y$ is the electroweak group responsible for weak and electromagnetic interactions. In addition to the gauge symmetry, there are other symmetries of the SM, such as baryon/lepton number conservation, (approximate) flavor symmetry, and so on. These can also be described by groups, called the global groups.
In this section, we will discuss the gauge groups and global groups in sequence to illustrate the applications of gauge theories and the group theories in particle physics.

\subsection{Gauge Group}

\paragraph{$SU(3)_C$}

Shortly after the quark model was proposed to explain the hadronic spectrum~\cite{Greenberg:1964pe,zweig1964_3}, a theoretical crisis emerged regarding the $\Omega^-$ baryon and the $\Delta^{++}$ resonance. According to the Pauli Exclusion Principle, a system of three identical fermions in the ground state (with symmetric spatial and spin wavefunctions) is forbidden. However, the $\Omega^-$ ($sss$) and $\Delta^{++}$ ($uuu$) both possess $J=3/2$, implying a totally symmetric spin-flavor-space state~\cite{struminsky1965magnetic}.

This paradox was resolved by introducing an additional internal degree of freedom~\cite{bogoliubov1965composite,Greenberg:1964pe}. This was later formalized as an $SU(3)_C$ gauge symmetry known as color~\cite{Han:1965pf}. By requiring the baryonic wavefunction to be a color singlet (fully anti-symmetric under $SU(3)_C$), the quarks could satisfy Fermi-Dirac statistics.
Subsequently, the interactions between these color charges were postulated to be mediated by an octet of vector gauge bosons, the gluons~\cite{Fritzsch:1973pi}.

The gluon fields forming $SU(3)_C$ octet $\mathbf{8}_c$ can be expanded by the 8 Gell-Mann matrices $\mathbf{\lambda}^a$
\begin{equation}
	G = \sum_{a=1}^8 G^a \mathbf{\lambda}^a \,.
\end{equation}
For quarks, the 3 colors, usually denoted by \textit{red, green and blue}, are collected in a triplet $\mathbf{3}$, also known as the fundamental representation,
\begin{equation}
	u = \left(\begin{array}{c}
u^r \\ u^g \\ u^b
\end{array}\right) \in \mathbf{3}_c\,,\quad d = \left(\begin{array}{c}
d^r \\ d^g \\ d^b
\end{array}\right) \in \mathbf{3}_c\,.
\end{equation}
To ensure the gauge symmetry, the normal derivative $\partial_\mu$ should be promoted to the covariant one,
\begin{align}
	D_\mu G_\nu &= \partial_\mu G_\nu + ig_c[G_\mu,G_\nu]\,, \\
	D_\mu q &= \partial_\mu q + ig_c G_\mu q\,,\quad q=u\,,d\,,
\end{align}
where $g_c$ is the coupling constant.
Putting them together, we obtain the $SU(3)_C$-invariant Lagrangian,
\begin{equation}
	\label{eq:qcdlagrangian}
	\mathcal{L} = -\frac{1}{2}\operatorname{Tr}G^{\mu\nu}G_{\mu\nu} + i\overline{u}\gamma^\mu D_\mu u + i\overline{d}\gamma^\mu D_\mu d\,.
\end{equation}
This resultant theory is known as quantum chromodynamics (QCD). Leptons do not participate in the strong interactions, t,us they are of the singlet representation of the $SU(3)_C$ group. In particular, we have omitted the mass terms of quarks in Eq.~\eqref{eq:qcdlagrangian}. Actually, the fermion masses are introduced in the SM by a scalar field known as the Higgs field, which will be discussed later.

\paragraph{$SU(2)_L\otimes U(1)_Y$}

The electroweak gauge group $SU(2)_L\otimes U(1)_Y$ is a unification of the weak and electromagnetic interactions, which was proposed before QCD in the 1960s by Glashow, Weinberg, and Salam~\cite{Glashow:1961tr,Weinberg:1967tq,Salam:1964ry}. To be concise, we consider only one flavor of fermions, and regard the right-handed neutrinos as not existing.

The electroweak group $SU(2)_L\otimes U(1)_Y$ is of dimension $3+1=4$, thus there are 4 gauge bosons, three of which form the adjoint representation of $SU(2)_L$, expanded by the generators $\mathbf{t}^i = \sigma^i/2$
\begin{equation}
	W = \sum_{i=1}^3W^i \mathbf{t}^i \,,
\end{equation}
and the other $B$ is neutral for $U(1)_Y$.
Since parity is not conserved in the weak interaction~\cite{Lee:1956qn,Wu:1957my}, the left- and right-handed fermions must be distinguished. Given the definition $\psi_{L/R}=P_{L/R}\psi$ with $P_{L/R}=(1\mp\gamma^5)/2$, the left-handed fermions are collected into $SU(2)_L$ doublets, while the right-handed fermions are $SU(2)_L$ singlets,
\begin{equation}
	q = \left(\begin{array}{c}
u_L \\ d_L
\end{array}\right) \in \mathbf{2}_L\,,\quad l = \left(\begin{array}{c}
\nu_{eL} \\ e_L
\end{array}\right) \in \mathbf{2}_L\,,\quad u_R\,,d_R\,,e_R \in \mathbf{1}_L\,.
\end{equation}
In particular, we have dropped the right-handed neutrino.
Besides, the Higgs field is needed to generate gauge boson masses, $H \in \mathbf{2}_L$, while we refer the interested readers to Ref.~\cite{Englert:1964et,Higgs:1964ia,Higgs:1964pj,Guralnik:1964eu}.
On the other hand, the Abelian gauge group $U(1)_Y$ corresponds to another conserving charge known as hypercharge $Y$.
It is determined by the Gell-Mann Nishijima formula~\cite{Nakano:1953zz,Nishijima:1955gxk,Gell-Mann:1956iqa} that
\begin{equation}
	Q = \frac{\sigma^3}{2} + \frac{Y}{2}\,,
\end{equation}
where $Q$ is the electric charge, $\sigma^3/2$ is the diagnal generators of the $SU(2)_L$ group.
As a consequence, we can obtain the hypercharges of the fermions and the Higgs field.
Similarly, the covariant derivatives are needed to ensure gauge symmetry,
\begin{align}
D_\mu\psi &= \partial_\mu \psi + ig_LW_\mu \psi + ig_Y Y B_\mu\psi\,,\quad \psi = q,l,u_R,d_R,e_R,H\,, \\
D_\mu W_\nu &= \partial_\mu W_\nu + ig_L[W_\mu,W_\nu] \,,\\
D_\mu B_\nu &= \partial_\mu B_\nu\,,
\end{align}
where $Y$ is the hypercharge of a specific field, $g_L$ and $g_Y$ are the coupling constants of $SU(2)_L$ and $U(1)_Y$ respectively. In particular, the covariant derivative of the Abelian gauge group $U(1)_Y$ is the same as the normal one.

In summary, we list the 3 gauge groups and the representation of the SM fields under them in Tab~\ref{tab:sm}. 
In particular, it can be verified that the SM is free of gauge anomaly.
The SM is a triumph of gauge theory and group theory in particle physics. After the discovery of the Higgs boson in 2012, the SM was verified completely. Up to now, the high-precision test of various processes is also aligned with the SM predictions. 


\begin{table}[ht]
\renewcommand{\arraystretch}{1.8}
	\begin{center}
		\begin{tabular}{|c|c|c|}
\hline
gauge group & gauge bosons & fermions \& scalar \\
\hline
$SU(3)_C$ & $G\in \mathbf{8}_c$ & $q\in \mathbf{3}_c\,,u_R\in\mathbf{3}_c\,, d_R\in\mathbf{3}_c$\\
\hline
\multirow{2}{*}{$SU(2)_L$} & \multirow{2}{*}{$W \in\mathbf{3}_L$} & $q\in \mathbf{1}_L\,,u_R\in\mathbf{2}_L\,,d_R\in\mathbf{1}_L$\\
& & $l\in\mathbf{2}_L\,,e_R\in\mathbf{1}_L\,,H\in\mathbf{2}_L$ \\
\hline
\multirow{2}{*}{$U(1)_Y$} & \multirow{2}{*}{$B \in 0_Y$} & $q\in \frac{1}{3}_Y\,, u_R\in\frac{4}{3}_Y\,,d_R\in-\frac{2}{3}_Y$ \\
& & $l\in -1_Y\,, e_R\in -2_Y\,, H\in \frac{1}{2}_Y$\\
\hline
		\end{tabular}
	\end{center}
	\label{tab:sm}
	\caption{The gauge groups, the gauge bosons, the fermions, and the Higgs field are listed. In particular, the nontrivial representations under the gauge groups are also presented.}
\end{table}

\subsection{Global Group}

To discuss the gauge group of the SM, we considered only one flavor of fermions. However, experiments imply there are 3 copies of the fermions, which share exactly the same gauge interactions, but have different masses, as shown in Tab.~\ref{tab:fermionmass}.

\begin{table}[hb]
\renewcommand{\arraystretch}{1.8}
\begin{center}
\begin{tabular}{cccc}
\hline
fermion & 1st flavor & 2nd flavor & 3rd flavor \\
\hline
\multirow{2}{*}{quark} & $m_u \approx 2.16\text{ MeV} $ & $m_c \approx 1.27\text{ GeV }$ & $m_t \approx 172.69\text{ GeV}$ \\
		       & $m_d \approx 4.67\text{ MeV} $ & $m_s \approx 93.4\text{ MeV}$ & $m_b \approx 4.18\text{GeV} $ \\
lepton & $m_e \approx 0.5109989500\text{ MeV}$ & $m_\mu \approx 105.6583755\text{ MeV } $ & $m_\tau = 1776.86\text{ MeV }$ \\
\hline
\end{tabular}
\end{center}
\label{tab:fermionmass}
\caption{The masses of the 3-flavor fermions in SM~\cite{ParticleDataGroup:2024cfk}.
}
\end{table}

Considering the 3-flavor fermions, we can write down the Yukawa interactions,
\begin{equation}
	\left(Y_u\right)_{pr}\overline{q}_p\tilde{H}u_R{}_r + \left(Y_d\right)_{pr}\overline{q}_p Hd_R{}_r + \left(Y_e\right)_{pr}\overline{l}_pH e_R{}_r+\text{h.c.}  \,,
\end{equation}
which are responsible for the fermion masses after the electroweak symmetry spontaneous breaking (EWSSB). Considering the 3 flavors, the Yukawa matrices are $3\times 3$ complex matrices, and we have written the flavor indices $p\,,r\,,\dots$ explicitly.

The Lagrangian implies that, in addition to assuming gauge symmetry, there exist other global symmetries. The first one is the flavor symmetry.
The fermion dynamics are diagonal in the flavor space, thus are invariant under the flavor group,
\begin{equation}
U(3)^5 = U(3)_q\otimes U(3)_u \otimes U(3)_d \otimes U(3)_l \otimes U(3)_e\,.
\end{equation}
This group is the maximal flavor group consistent with the SM gauge group~\cite{Chivukula:1987py,Gerard1983FermionMS,Sun:2025axx}. Nevertheless, the flavor group, opposite to the gauge group, is not exact. The $U(3)^5$ group is broken by the Yukawa matrices. On the other hand, $U(3)^5$ is not broken completely, and the unbroken subgroup can be identified by adopting a specific flavor basis. For example, in the down basis, the Yukawa matrices are parameterized as
\begin{equation}
	Y_u = V_{CKM}^\dagger\text{diag}(y_u,y_c,y_t)\,,\quad Y_d=\text{diag}(y_d,y_s,y_b)\,,\quad Y_e = \text{diag}(y_e\,,y_\mu\,,y_\tau)\,,
\end{equation}
with the CKM-matrix $V_{CKM}$~\cite{Cabibbo:1963yz,Kobayashi:1973fv}. The unbroken subgroup is Abelian, 
\begin{equation}
	U(3)^5 \rightarrow U(1)^4 = U(1)_B \otimes U(1)_L \otimes U(1)_{e-\mu} \otimes U(1)_{\tau-\mu}\,,
\end{equation}
called rephasing symmetry.
$U(1)_B$ ensures baryon number conservation, and $U(1)_L \otimes U(1)_{e-\mu} \otimes U(1)_{\tau-\mu}$ ensures the lepton number conservations of the three flavors.
In the SM, the flavor symmetry $U(3)^5$ is broken, but the rephasing symmetry $U(1)^4$ is exact.

As discussed before, discrete symmetries such as parity and charge conjugation can also be complemented by group theory. Given the multiple flavors of the fermions in SM, the combination of parity and charge conjugation, called $CP$ transformation, is closely related to the flavor symmetry, since on one hand, rephasing symmetry could eliminate CP-violating phases, on the other hand, rephasing symmetry is determined by the flavor symmetry~\cite{Sun:2025axx}. For example, the $U(3)^5$ flavor group reduces to the $U(1)^4$ rephasing group, under which the only CP-violating phase emerges in the CKM-matrix~\cite{Jarlskog:1985ht,Jarlskog:1985cw,Bernabeu:1986fc}.

\paragraph{Custodial Symmetry}


Unlike the flavor symmetry, custodial symmetry is an important global symmetry of not fermions but the Higgs sector~\cite{Longhitano:1980tm,Longhitano:1980iz,Appelquist:1980vg}. Considering the SM Lagrangian of Higgs,
\begin{equation}
	\label{eq:higgslagrangian}
	\mathcal{L}_{\text{Higgs}} = (D_\mu H^\dagger)(D^\mu H) - V(HH^\dagger) = \frac{1}{2}(\partial_\mu\phi)\cdot(\partial^\mu\phi) - V(\phi\cdot\phi)\,,
\end{equation}
where we have expressed the Lagrangian by the four real fields $\vec{\phi}=(\phi^1,\phi^2,\phi^3,\phi^4)$ in the complex Higgs doublet,
\begin{equation}
	H = \frac{1}{\sqrt 2}\left( \begin{array}{c}
	\phi^2+i\phi^1\\ 
	\phi^4-i\phi^3
	\end{array}
	\right)\,.
\end{equation}
The Higgs Lagrangian in Eq.~\eqref{eq:higgslagrangian} is invariant under the $O(4)$ group acting on $\vec{\phi}$ as $\vec{\phi}\rightarrow O\vec{\phi}\,,O\in O(4)$. This $O(4)$ symmetry is called the custodial symmetry, which is an exact global group of the Higgs sector.

Nevertheless, the custodial group is broken not only spontaneously but also explicitly. To describe the spontaneous breaking, we utilize the local equivalence $SO(4) \cong SU(2)_L\times SU(2)_R$, and define a Higgs field as 
\begin{equation}
	\Sigma = (\tilde{H},H)\,,
\end{equation}
with the dual Higgs field $\tilde{H} = i\sigma_2 H$. $\Sigma$ is covariant under the custodial symmetry
\begin{equation}
\Sigma \rightarrow L\Sigma R^\dagger\,,\quad (L,R)\in SU(2)_L\times SU(2)_R\,,
\end{equation}
and the Lagrangian Eq.~\eqref{eq:higgslagrangian} becomes
\begin{equation}
	\mathcal{L}_{\text{Higgs}} = \frac{1}{2}\text{tr}\left[(\partial_\mu \Sigma)^\dagger(\partial^\mu\Sigma)\right] - V(\text{tr}(\Sigma^\dagger\Sigma))\,.
\end{equation}
The Higgs vacuum expectation value (VEV) raises the spontaneous breaking $SU(2)_L\times SU(2)_R\rightarrow SU(2)_V$, where $SU(2)_V$ is the diagonal subgroup, $SU(2)_V=\{(L,R)\in SU(2)_L\times SU(2)_R|L=R\}$. On the other hand, the custodial symmetry is explicitly broken by two aspects. One is the Yukawa terms, since the masses of the fermions in the same $SU(2)_L$ doublets are different, the other is the gauge interactions, since the $SU(2)_L$ is gauged completely, while only the third component of $SU(2)_R$ is gauged.

Although the global symmetries are usually not exact, they are also important in the gauge theory. On the one hand, the global symmetry also determines the dynamics. For example, the chiral symmetry $SU(3)_L\times SU(3)_R$ is the flavor symmetry of the low-energy QCD, and it determines the particle spectrum and the interactions of light hadrons~\cite{Weinberg:1978kz,Gasser:1983yg,Gasser:1984gg}. 
On the other hand, the global symmetries differ in different gauges. For example, the baryon-number and lepton-number conserving groups $U(1)_B$ and $U(1)_L$ are exact in the SM, but break in a theory of a larger gauge group, such as the grand unified theory based on the $SU(5)$ gauge group~\cite{Georgi:1974sy}.

\section{Conclusions}
\label{sec:conclusions}


The journey from the abstract symmetries of spacetime to the intricate internal symmetries of the Standard Model stands as a monumental testament to the power of fundamental principles in theoretical physics. In this grand synthesis, group theory provides the immutable mathematical skeleton, offering the language to classify particles and restrict their possible interactions. Complementing this, the gauge principle provides the dynamical flesh, weaving the very fabric of forces from the fundamental thread of local invariance. This is exquisitely demonstrated in Quantum Electrodynamics, the physical realization of the U(1) gauge theory. The true power of this principle was unlocked with its generalization to non-Abelian groups, where the requirement of local symmetry not only dictates the existence of interactions but also leads to the profound phenomenon of self-interacting force carriers. The culmination of this paradigm is the Standard Model, a Yang-Mills theory based on the gauge group $SU(3)_C \times SU(2)_L \times U(1)_Y$, which unifies the electroweak and strong dynamics into a single, predictive framework that describes almost all known fundamental phenomena.

Yet, the exploration is far from complete. Today, physicists vigorously employ these very tools—symmetry and group theory—as their guiding principles to probe the physics that lies beyond the Standard Model. This endeavor proceeds along two complementary paths: the top-down approach, which seeks a more fundamental underlying theory, and the bottom-up approach, which systematically parameterizes our ignorance.

The top-down approach is driven by the quest for a more complete and unified theory. Key directions involve strategic extensions of the symmetry structures that have proven so successful:
\begin{itemize}
    \item Extension of Gauge Groups: This includes models like the left-right symmetric model $SU(2)_L \times SU(2)_R \times U(1)$~\cite{Mohapatra:1974hk}, the Pati-Salam model~\cite{Pati:1974yy}, and ultimately Grand Unified Theories (GUTs) such as SU(5) and SO(10)~\cite{Georgi:1974sy,Georgi:1975tx,Fritzsch:1974nn}, which aim to merge the strong and electroweak forces into a single simple group.

    \item Extension of Global Symmetries: Theories like the composite Higgs model~\cite{Kaplan:1984sm,Georgi:1984af,Dugan:1985gm,Arkani-Hamed:2002ikv,Arkani-Hamed:2002iiv,Agashe:2004rs,Agashe:2005dk,Agashe:2006at,Contino:2006qr} postulate that the Higgs boson is not elementary, but part of a larger multiplet, requiring a bigger global symmetry group for its description.

    \item Extension of Spacetime Symmetry: The Coleman-Mandula theorem~\cite{Coleman:1967ad} severely restricts the possible unions of internal and spacetime symmetries, but it allows for one profound extension: supersymmetry~\cite{Wess:1974tw,Salam:1974yz}. This symmetry relates bosons and fermions, placing them within supermultiplets, and offers solutions to the Higgs hierarchy problem and a pathway to dark matter.

    \item Extension of Spacetime Dimension: Models with extra spatial dimensions, such as the ADD~\cite{Arkani-Hamed:1998jmv} and Randall-Sundrum scenarios~\cite{Randall:1999ee,Randall:1999vf}, propose that the fundamental scale of gravity is much lower, with the apparent weakness of gravity arising from the geometry of the hidden dimensions.
\end{itemize}

In parallel, the bottom-up approach is encapsulated by the Standard Model Effective Field Theory (SMEFT)~\cite{Isidori:2023pyp}
. This framework assumes that the Standard Model fields are the relevant degrees of freedom at low energies and systematically constructs all possible interaction terms consistent with the known spacetime and gauge symmetries~\cite{Weinberg:1978kz}.
The resulting Lagrangian is organized as an expansion in operator dimensions, which has been written up to dimension 9~\cite{Buchmuller:1985jz,Grzadkowski:2010es,Lehman:2014jma,Liao:2016hru,Li:2020gnx,Li:2020xlh,Harlander:2023psl,Henning:2015alf}. To manage the proliferation of higher-dimensional operators, approximate global symmetries—such as CP, flavor~\cite{Sun:2025axx,Remmen:2019cyz,Durieux:2024zrg}, baryon number and lepton number~\cite{Weinberg:2020zwl,Fonseca:2020xlj}, and custodial symmetry~\cite{Kribs:2020jgn}—are imposed to organize and constrain the theory, providing a model-independent window into new physics, parameterized by Higgs Effective Field Theory (HEFT)~\cite{Alonso:2012px,Buchalla:2013rka,Brivio:2016fzo,Sun:2022ssa,Sun:2022snw,Buchalla:2013eza,Buchalla:2017jlu}. At lower-energy scale, the Low-energy Effective Field Theory (LEFT)~\cite{Jenkins:2017jig,Jenkins:2017dyc,Liao:2020zyx,Li:2020tsi} describes new physics after the electroweak gauge symmetry is spontaneously broken, and the chiral perturbation theory~\cite{Weinberg:1966fm,Gasser:1984gg,Ebertshauser:2001nj,Li:2024ghg,Fettes:2000gb,Song:2024fae} describes the hadronic interactions after the chiral symmetry is spontaneously broken.

Looking further ahead, the frontier of theoretical physics continues to be shaped by novel symmetry concepts. 
The concept of generalized symmetry, including higher-form and non-invertible symmetries, has attracted growing interest in both high-energy physics and condensed matter physics since the recent seminal paper ~\cite{Gaiotto:2014kfa}, though development of the relevant concept dates back to early 90s~\cite{Alford:1990fc,Alford:1991vr,Alford:1992yx,Bucher:1991bc}. See Ref.~\cite{Cordova:2022ruw,Gomes:2023ahz,Brennan:2023mmt,Schafer-Nameki:2023jdn} for reviews.
In this framework, symmetry operations are associated with topological operators, and the corresponding charged objects are generalized to higher-dimensional extended objects rather than local fields.
For instance, the center symmetry of the $SU(N)$ gauge group can be interpreted as a 1-form electric (magnetic) symmetry acting on the Wilson (’t Hooft) lines~\cite{Gaiotto:2014kfa,Tong:2017oea,Bolognesi:2019fej}. These 1-form symmetries are genuine global symmetries and hence physical. They can be gauged or broken and can also possess anomaly.
In physical applications, the breaking of such electric 1-form symmetries provides a useful criterion for distinguishing the confining and deconfining phases of a gauge theory.
Gauging a 1-form symmetry introduces a 2-form gauge field, thereby modifying the global structure of the gauge group and constraining the spectrum of line operators.

Moreover, anomaly matching for higher-form symmetries serves as a valuable tool for probing phenomena such as chiral symmetry breaking in strongly coupled gauge theories~\cite{Shimizu:2017asf,Bolognesi:2021yni}.
The notion of non-invertible symmetry, as the name suggests, refers to symmetry operations that lack an inverse. The associated topological operators form fusion categories rather than groups. Applications of non-invertible symmetries in particle physics remain an active area of research~\cite{Choi:2022jqy,Yokokura:2022alv,Putrov:2023jqi,Choi:2023pdp,Cordova:2022fhg,Cordova:2023her,Kobayashi:2024cvp,Cao:2024lwg,Kobayashi:2025znw,Suzuki:2025oov}.
Similar the higher form symmetry they can also be gauge, or one can match their 't Hooft anomaly leaving constraint on the renoramlization group flow of the system~\cite{Chang:2018iay}.
The selection rule in the associated mathematical structure -- fusion-algebra are also used to suppress specific couplings in model building and their phenomenology~\cite{Cordova:2022fhg,Hidaka:2024kfx,Kobayashi:2024cvp,Kobayashi:2025znw,Suzuki:2025oov,Kobayashi:2025ldi,Kobayashi:2025cwx,Kobayashi:2025lar,Suzuki:2025kxz,Okada:2025adm,Nakai:2025thw}.

In conclusion, the relentless pursuit of symmetry, articulated through group theory and enacted through the gauge principle, has been the most reliable compass in fundamental physics. From the consolidation of the known forces in the Standard Model to the bold ventures beyond it, this principle remains our primary guide. The ultimate goal endures: to discover a unified description of all forces, including gravity, in the hope that an even deeper, more encompassing symmetry principle—perhaps one still waiting to be conceived—ultimately governs the cosmos.

\begin{ack}[Acknowledgments]%
This work is supported by the
National Science Foundation of China under Grants No. 12347105, No. 12375099, No. 12447101 and No.1250050417, and the National Key Research and Development Program of China Grant No. 2020YFC2201501,
No. 2021YFA0718304, and the start-up funding of Sun Yat-Sen University under grant number 74130-122550.
M.-L.X. is supported by the National Natural Science Foundation of China (Grant No.12405123), Fundamental Research Funds for the Central Universities, Sun Yat-sen University (Grant No.25hytd001), Shenzhen Science and Technology Program (Grant No.JCYJ20240813150911015). 
\end{ack}


\seealso{The pedagogical introductions to group theory and gauge theory can be found, for example, in the following books and lecture notes: \cite{Tung:1985iqd,Zee:2016fuk,ma2007group,Cheng:1984vwu,AitchisonHey2012,Pokorski:1987ed,Weinberg:1996kr,Schwartz:2014sze,Donoghue:1992dd,Burgess:2006hbd}.
}


\bibliographystyle{JHEP} 
\bibliography{reference}

\end{document}